\documentclass[a4paper,fleqn,usenatbib]{mnras}

\usepackage{amsmath}	
\usepackage{mathptmx}
\usepackage{txfonts}

\usepackage[T1]{fontenc}
\usepackage{ae,aecompl}

\usepackage{graphicx}	
\usepackage{pdflscape}  
\usepackage{caption}

\title[High-resolution optical spectroscopy of RS Ophiuchi during 2008 -- 2009]{High-resolution optical spectroscopy of RS Ophiuchi during 2008 -- 2009}

\author[A. Somero, P. Hakala and G. Wynn]{A. Somero$^{1}$\thanks{E-mail:
aunsom@utu.fi}, P. Hakala$^{2}$ and G.A. Wynn$^{3}$
\\
$^{1}$Tuorla Observatory, Department of Physics and Astronomy, University of Turku, V\"{a}is\"{a}l\"{a}ntie 20, FI-21500 Piikki\"{o}, Finland\\
$^{2}$Finnish Centre for Astronomy with ESO (FINCA), University of Turku, V\"{a}is\"{a}l\"{a}ntie 20, FI-21500 Piikki\"{o}, Finland\\
$^{3}$Department of Physics \& Astronomy, University of Leicester, Leicester LE1 7RH, UK}

\date{Accepted 2016 October 3. Received 2016 September 14; in original form 2016 August 12}

\pubyear{2016}

\begin{document}
\label{firstpage}
\pagerange{\pageref{firstpage}--\pageref{lastpage}}
\maketitle

\begin{abstract}
RS Ophiuchi is a symbiotic variable and a recurrent nova.
We have monitored it with the Nordic Optical Telescope and obtained 30 high resolution (R=46 000) optical spectra over one orbital cycle during quiescence.
To our knowledge this is the best-sampled high resolution spectroscopic dataset of RS Oph over one orbital period.
We do not detect any direct signatures of an accretion disc such as double peaked emission lines, but many line profiles are complex consisting of superimposed emission and absorption components.
We measure the spin of the red giant and conclude that it is tidally locked to the binary orbit.
We observe Na I absorption features, probably arising from the circumbinary medium, that has been shaped by previous recurrent nova outbursts.
We do not detect any intrinsic polarisation in the optical wavelengths.
\end{abstract}




\section{Introduction}

RS Ophiuchi (RS Oph) is a symbiotic binary and a recurrent nova (RN) composed of a white dwarf (WD) that accretes matter from an M type giant companion star \citep{dobrzycka94, kenyon86}.
The system is embedded in the stellar wind of the red giant. 
The orbital period of RS Oph is 453.6 days \citep{brandi}.

RS Oph exhibits recurrent outbursts approximately every 20 years and so far six outbursts have been recorded in 1898, 1933, 1958, 1967, 1985 and 2006 \citep{brandi}.
\citet{schaefer2010} also argues that two outbursts were missed in 1907 and 1945.
The recurrent nova outbursts are thought to be the result of a thermonuclear runaway on the WD.
Because of the short recurrence time of the outbursts the WD mass has been estimated to be close to Chandrasekhar limit and thus it is considered a strong candidate for a future type Ia supernova explosion \citep{hernanz, starrfield}.
However, \citet{king} and \citet{alexander} suggest that the outbursts of RS Oph could be due to an accretion disc instability, analogous to dwarf nova outbursts.

It is unclear whether the accretion process in RS Oph occurs via Roche lobe overflow or from the stellar wind capture.
However, \citet{schaefer} has calculated that in order to accrete enough mass between the outbursts the red giant should fill its Roche lobe.

The outburst in 2006 was extensively observed across the electromagnetic spectrum and a plethora of articles have been published about it \citep[see e.g.,][]{rsoph2006conf}.
However, the previously published observations obtained during quiescence often consist of sparse data points or data obtained over several orbital periods \citep[e.g.][]{zamanov, brandi, patat}.

The motivation of our study is to find out whether the mass transfer during quiescence occurs via Roche lobe overflow or stellar wind capture.
A better understanding of the system behaviour during quiescence will help determine the nature of the accretion process and the associated outbursts \citep{alexander}.
In this paper we present the data and report on the results of our monitoring campaign of RS Oph in quiescence.

In Section 2 we describe the observations, followed by the data analysis in Sections 3 and 4.
Finally, we present the discussion and conclusions.

\section[]{Observations}
\subsection{FIES}
We observed RS Oph with the Fiber-Fed Echelle Spectrograph (FIES) on the 2.5-m Nordic Optical Telescope (NOT) \citep{Telting2014} at La Palma during 2008 and 2009.
The observations were conducted as a monitoring campaign.
The plan was to take one spectrum roughly every two weeks, but this was hampered by bad weather and the visibility of the target.
In total we collected 30 spectra over one orbital period of RS Oph (see Table \ref{Tab:ObsLog} for the observing log).
To our knowledge, this is so far the best-sampled high-resolution spectroscopic dataset of RS Oph over one whole orbital period.

The spectra were obtained using the medium-resolution fiber (R=46000, aperture of 1.3\arcsec) and an exposure time of 1200 s.
A ThAr arc lamp spectrum was exposed immediately before and after the science spectrum.

Before the data reduction, cosmic rays were removed by median filtering the frames. 
The data were reduced with FIEStool\footnote{http://www.not.iac.es/instruments/fies/fiestool/FIEStool.html}, a FIES data reduction pipeline provided by the NOT.
Heliocentric correction was applied to the spectra before further processing.
In total 78 orders were extracted from the spectra corresponding to the wavelength range of 3635 -- 7271 \AA.
The blue end of the spectra are noisy due to the short exposure time relative to the low UV sensitivity of the spectrograph.

\begin{figure*}
\begin{center}
\includegraphics[width=0.3\textwidth, angle=90]{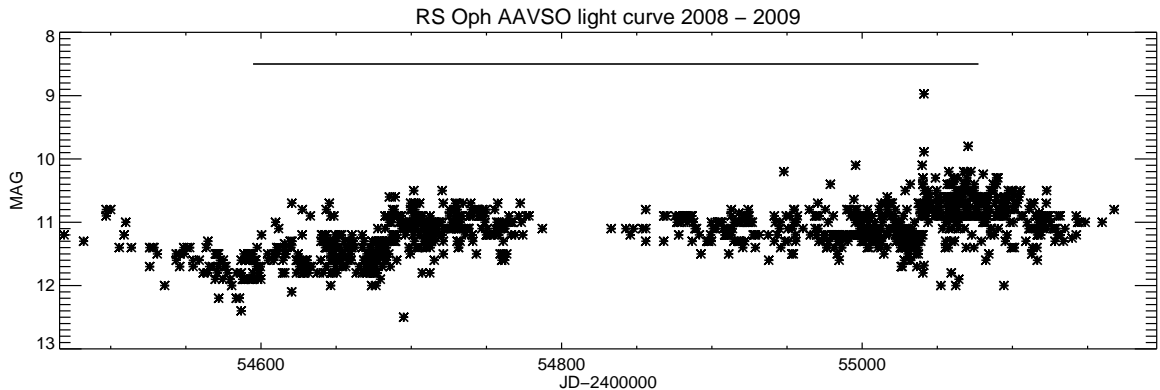}
\caption{Visual band light curve of RS Oph by AAVSO covering our observations which are marked with the horizontal line. The plot includes only the values given by AAVSO, no errors.}
\label{pic:aavso}
\end{center}
\end{figure*}

Figure \ref{pic:aavso} 
shows the long term light curve of RS Oph covering the time window of our observations, obtained from the American Association of Variable Star Observers (AAVSO).

\begin{table}
 \centering
 \begin{minipage}{75mm}
  \caption{Log of observations. Date refers to the date of the start of night on La Palma. S/N of the spectra varies between 17 and 40 calculated around 5900 \AA.}
  \label{Tab:ObsLog}
  \begin{tabular}{@{}llcc@{}}
  \hline
\# &Date & HJD at mid exposure & Orbital phase\footnote{According to ephemeris by \citet{brandi}} \\
 & & $+ 2 400 000$ & \\
 \hline
1 & 2008-05-05 & 54592.7031 & 21.05 \\
2 & 2008-05-11 & 54598.6303 & 21.07 \\
3 & 2008-05-31 & 54618.6522 & 21.11 \\
4 & 2008-06-30\footnote{exposure time 1127.204 s} & 54648.5486 & 21.18 \\
5 & 2008-07-15 & 54663.4260 & 21.21 \\
6 & 2008-07-28 & 54676.4075 & 21.24 \\
7 & 2008-08-13 & 54692.4320 & 21.27 \\
8 & 2008-08-27 & 54706.3629 & 21.30 \\
9 & 2008-09-14\footnote{clouds, exposure time 1800 s} & 54724.3543 & 21.34 \\
10 & 2008-09-30 & 54740.3719 & 21.38 \\
11 & 2008-10-12 & 54752.3510 & 21.40 \\
12 & 2008-10-21 & 54761.3197 & 21.42 \\
13 & 2008-11-11\footnote{clouds and high airmass} & 54782.3171 & 21.47 \\
14 & 2008-11-16 & 54787.3067 & 21.48 \\
15 & 2009-01-21 & 54853.7976 & 21.63 \\
16 & 2009-01-23 & 54855.7972 & 21.63 \\
17 & 2009-02-02\footnote{lower hatch closed (wind)} & 54865.7794 & 21.65 \\
18 & 2009-02-11 & 54874.7782 & 21.67 \\
19 & 2009-03-19 & 54910.6695 & 21.75 \\
20 & 2009-03-30 & 54921.6968 & 21.78 \\
21 & 2009-04-16 & 54938.7388 & 21.81 \\
22 & 2009-04-29 & 54951.6564 & 21.84 \\
23 & 2009-05-17 & 54969.7196 & 21.88 \\
24 & 2009-06-05 & 54988.5438 & 21.92 \\
25 & 2009-06-14 & 54997.6508 & 21.94 \\
26 & 2009-06-24 & 55007.6818 & 21.97 \\
27 & 2009-07-10 & 55023.5298 & 22.00 \\
28 & 2009-07-25 & 55038.4176 & 22.03 \\
29 & 2009-08-16 & 55060.4991 & 22.08 \\
30 & 2009-09-02 & 55077.3859 & 22.12 \\
\hline
\end{tabular}
\end{minipage}
\end{table}

\subsection{ALFOSC}
In addition, we obtained one data set of linear spectropolarimetry of RS Oph.
The observation was done also with the NOT but using the Andalucia Faint Object Spectrograph (ALFOSC) on 13 August 2008.
The data were obained with a $\lambda/2$ retarder plate together with a 1.2\arcsec polarimetric slitlet on a Calcite and a grism (\#4: 3200 -- 9100\AA, R $\sim $ 236) with an exposure time of 300 s.
The retarder plate was rotated in a sequence and exposures were taken with 4 different angles (0$^{\circ}$, 22.5$^{\circ}$, 45$^{\circ}$, 67.5$^{\circ}$).
We also observed one zero polarisation standard (HD2121311, exposure time 80 s) and high polarisation standard (HD204827, 60 s).
The standard stars were observed on 18 August 2008 with the same instrument setup.

The spectra were bias corrected and optimally extracted in IRAF\footnote{IRAF    is    distributed    by    the    National    Optical    Astronomy
Observatories,  which  are operated by  the Association of Universities
for Research in Astronomy, Inc., under cooperative agreement with the
National Science Foundation}.
Wavelength calibration and further processing were done in MOLLY, a software package developed by Tom Marsh\footnote{http://deneb.astro.warwick.ac.uk/phsaap/software/}.
The reduced spectra cover a wavelength range of 3200 -- 9000 \AA, however we cut the spectra at 8200 \AA\ due to noise in the red end.
The spectra were binned to 146 data points (each bin corresponding to 30 \AA) and zero-corrected before the degree of polarisation was calculated.
The errors were estimated from the scatter of the continuum.

\section{High-resolution spectra}

\begin{figure*}
\begin{center}
\includegraphics[width=1.0\textwidth, angle=90]{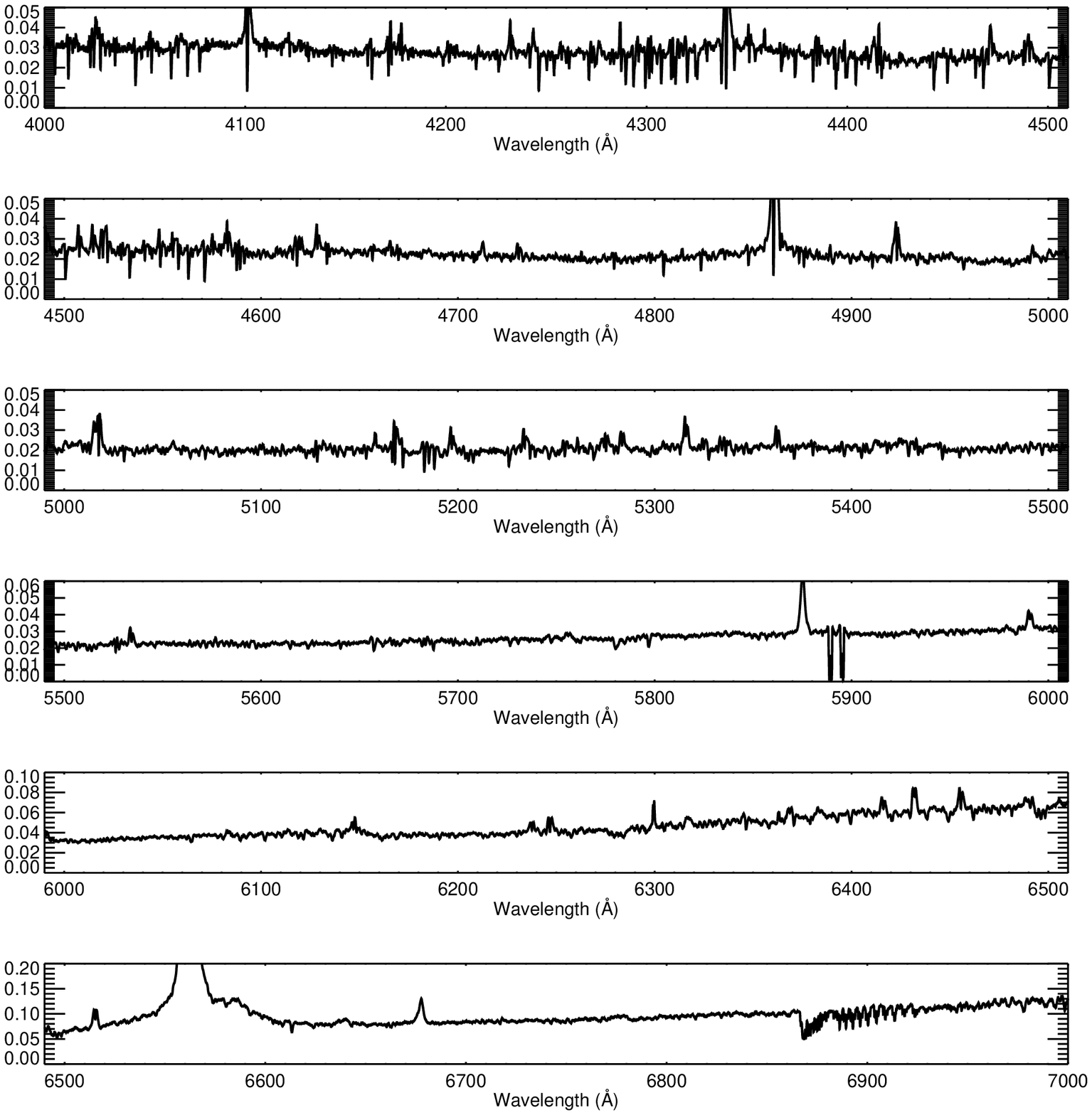}
\caption{The average spectrum of RS Oph. The absorption features from around 6850 \AA\ onwards are due to the atmosphere.}
\label{pic:ave}
\end{center}
\end{figure*}

The average of all RS Oph FIES spectra is shown in Fig. \ref{pic:ave}.
The spectrum shows emission features of He I (5875 and 7065 \AA), complex features of H (6562, 4861, 4340 and 4101 \AA) that are a combination of emission and absorption, and features of Fe II (4233, 4924, 5018A, 5196, 5197, 5235, 5276, 5217, 5365, 5535 and 5991 \AA), many of them from the red giant, resembling the quiescent spectrum described by \citet{anupamamikolajewska}.
We do not see signs of He II 4686 \AA\ nor O III emission, which are detected in some other symbiotic binaries \citep[see e.g.][]{ikedatamura} or previously in quiesence spectra of RS Oph before the 2006 outburst \citep{zamanov}.

RS Oph also shows an unusually strong Li I 6707 \AA\ absorption line \citep{wallerstein, brandi} which is also present in our spectra.

\subsection{Line profiles}

\begin{figure*}
\begin{center}
\includegraphics[width=0.3\textwidth, angle=90]{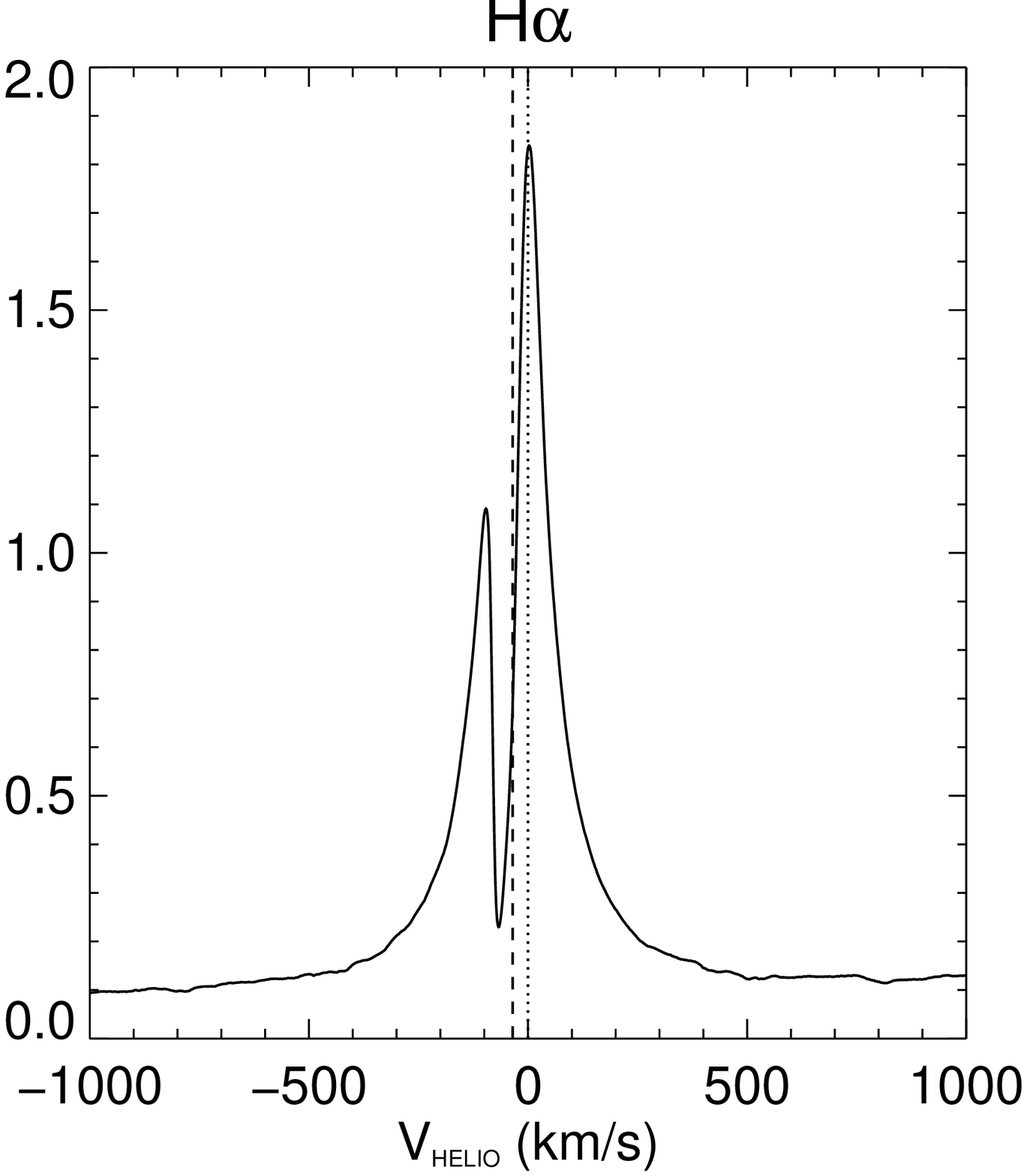}
\includegraphics[width=0.3\textwidth, angle=90]{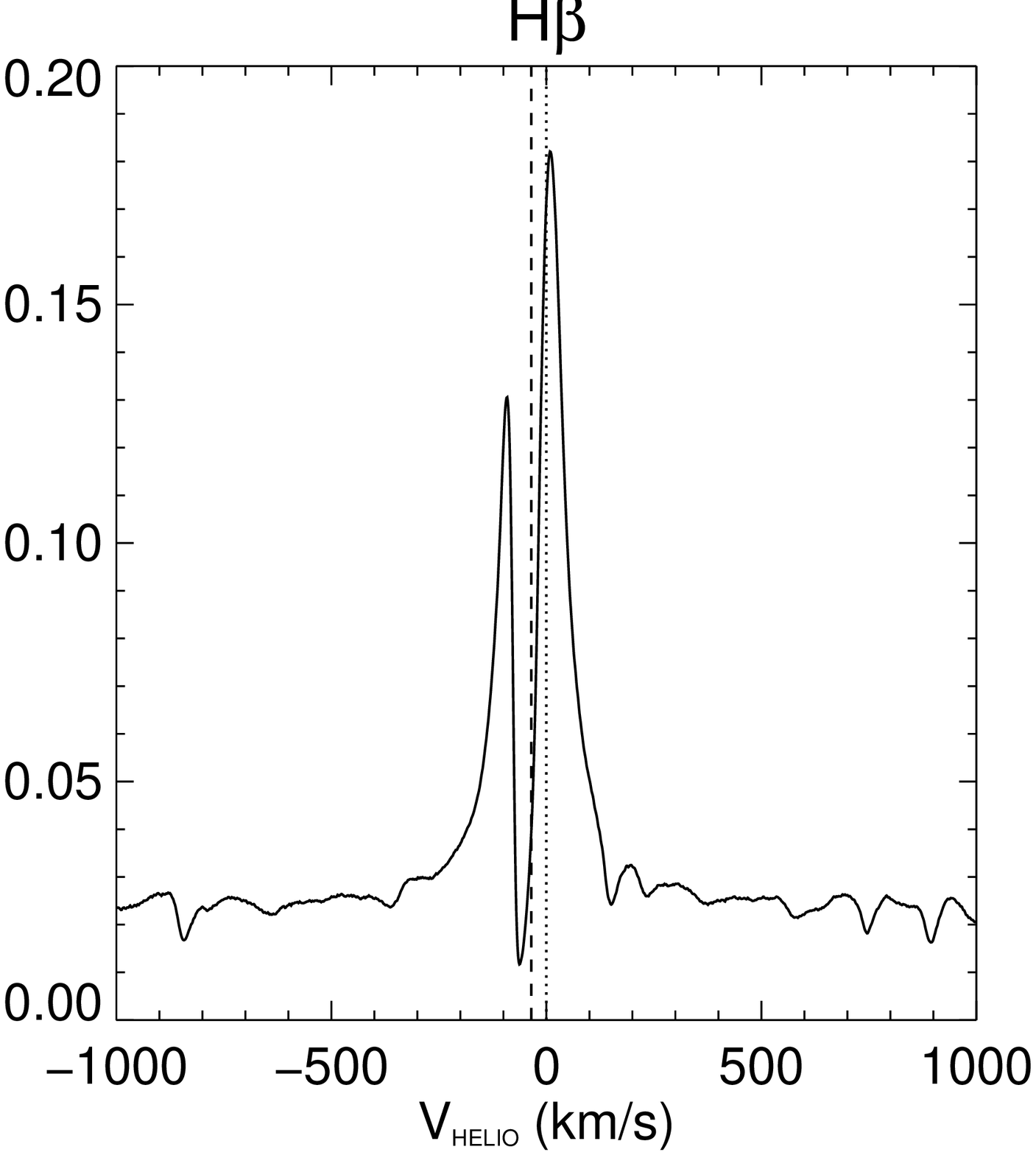}
\includegraphics[width=0.3\textwidth, angle=90]{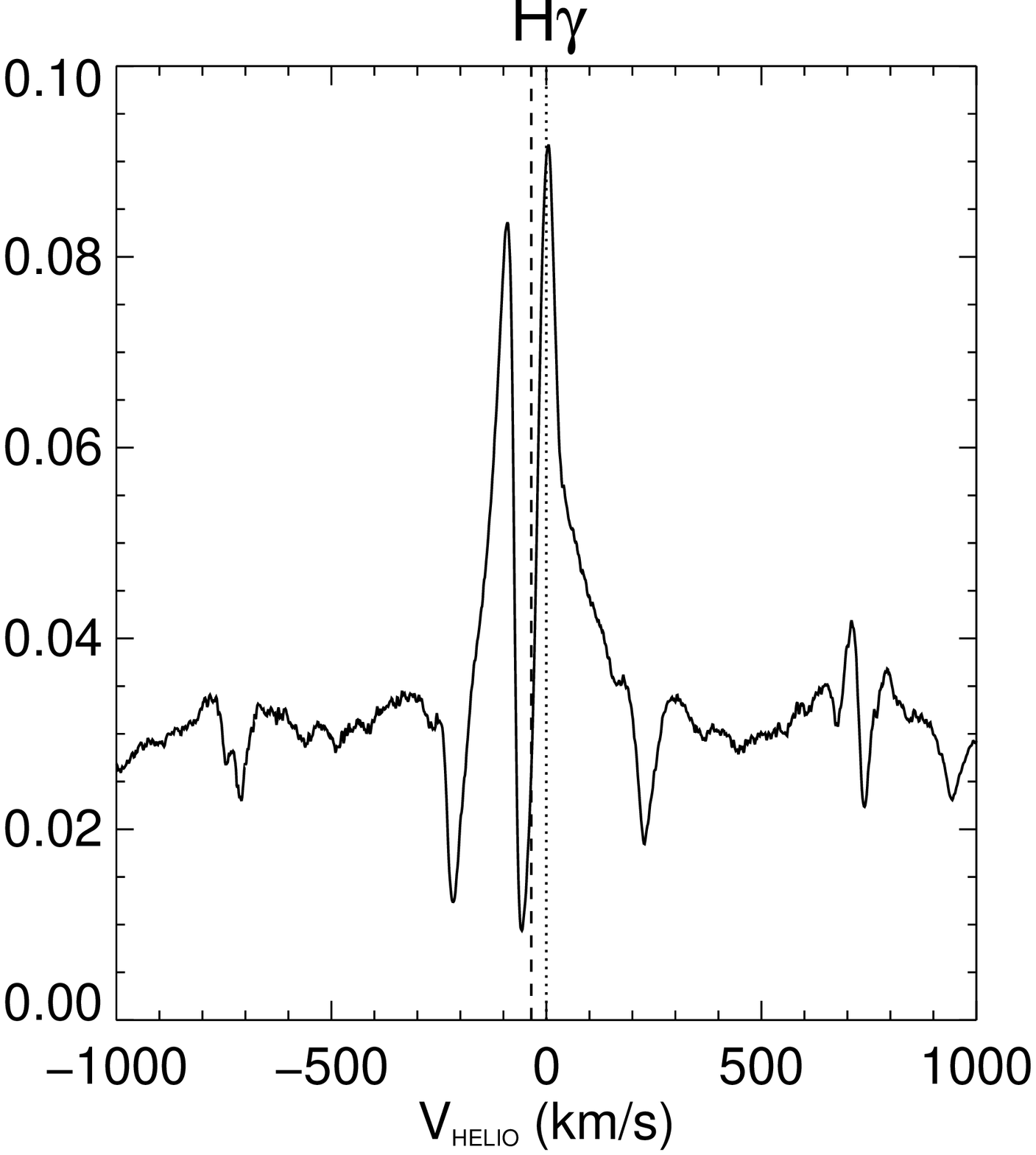}
\includegraphics[width=0.3\textwidth, angle=90]{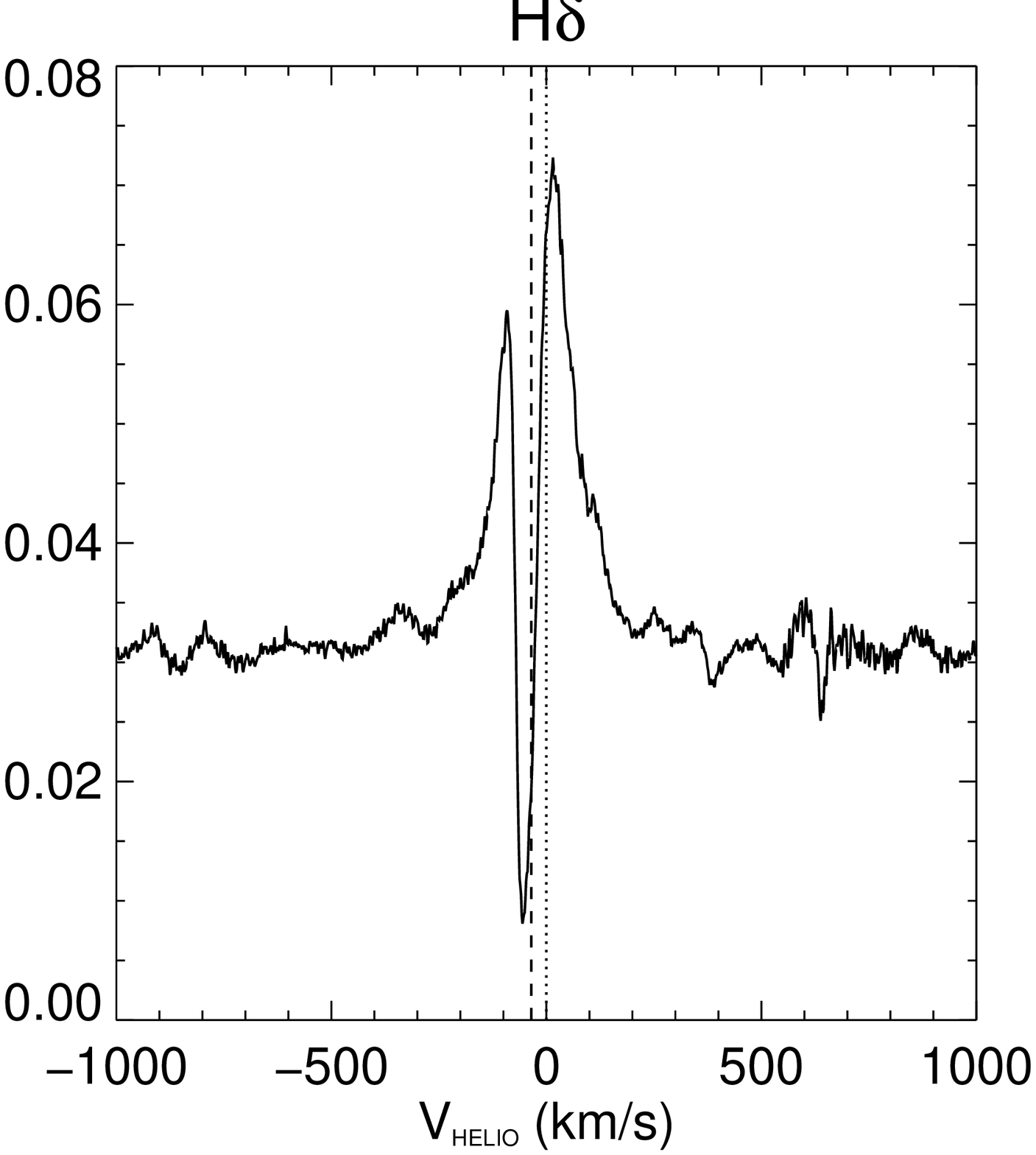}
\includegraphics[width=0.3\textwidth, angle=90]{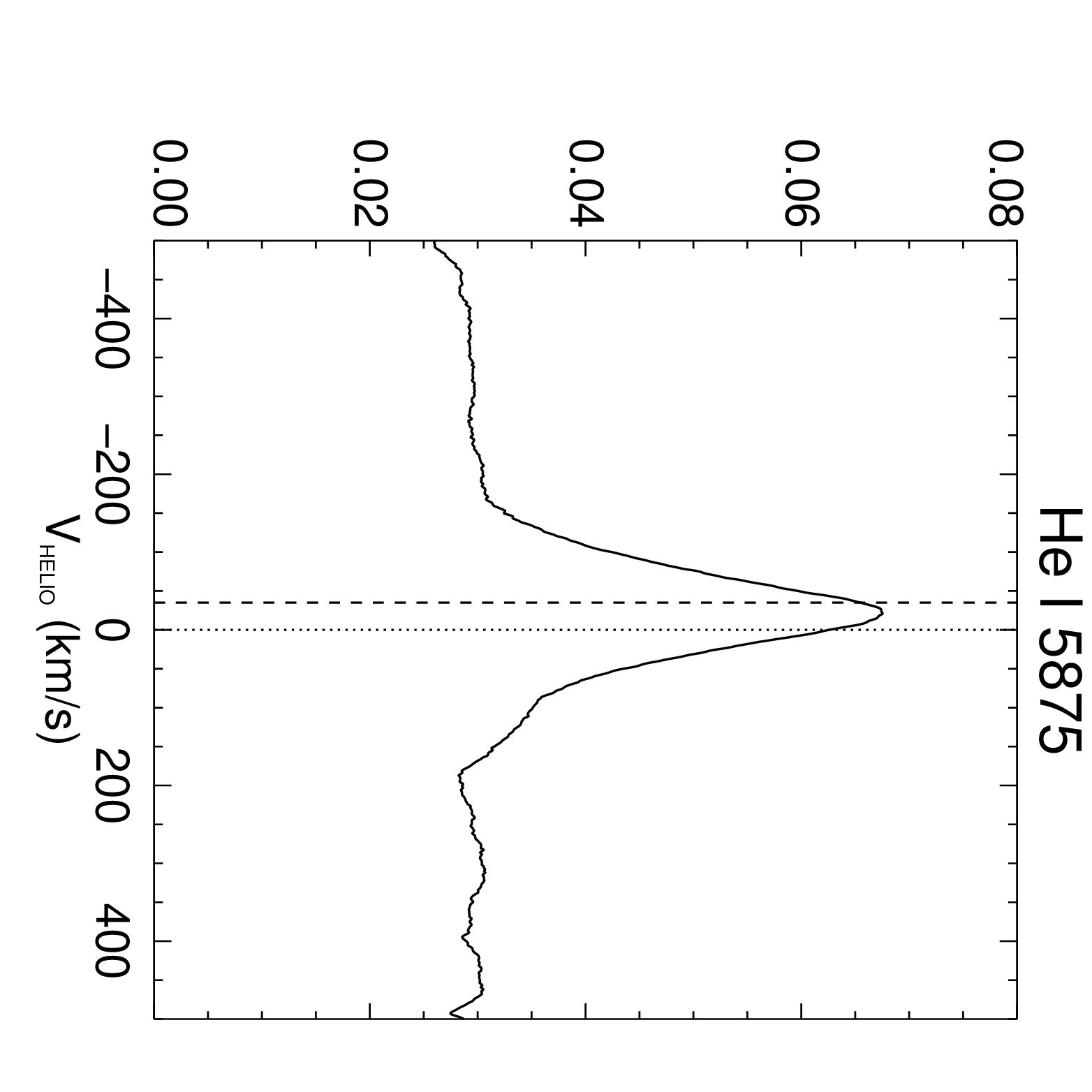}
\includegraphics[width=0.3\textwidth, angle=90]{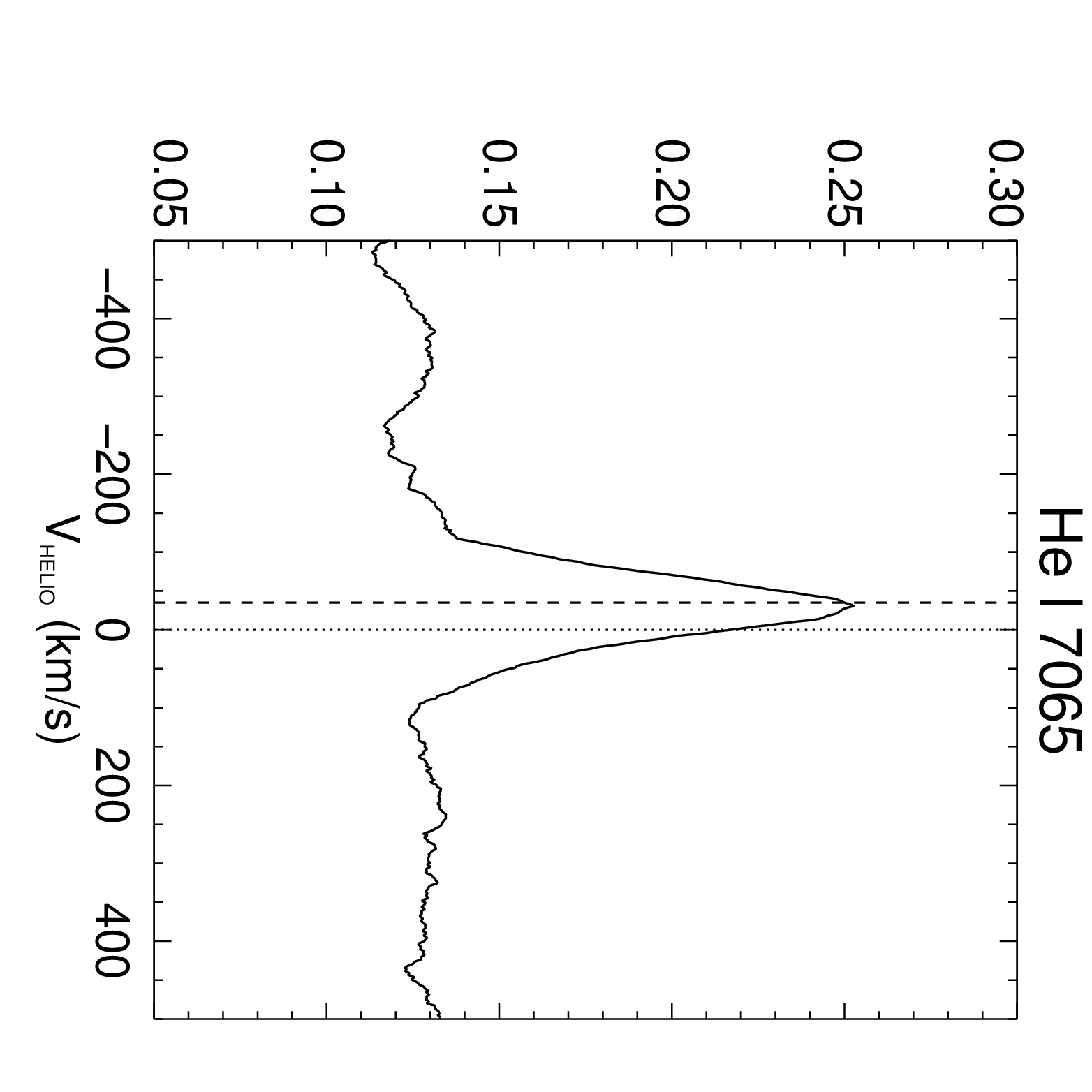}
\includegraphics[width=0.3\textwidth, angle=90]{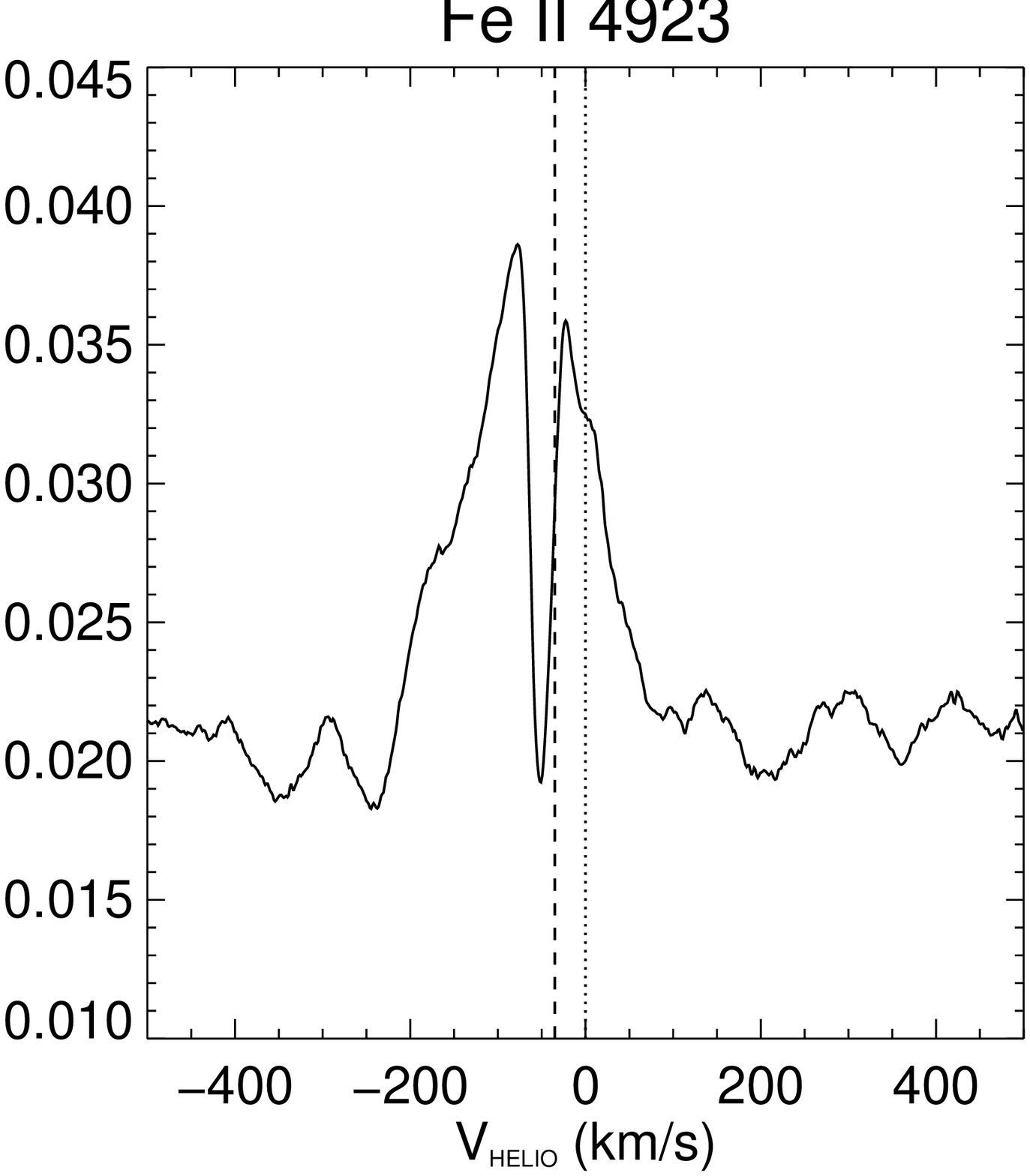}
\includegraphics[width=0.3\textwidth, angle=90]{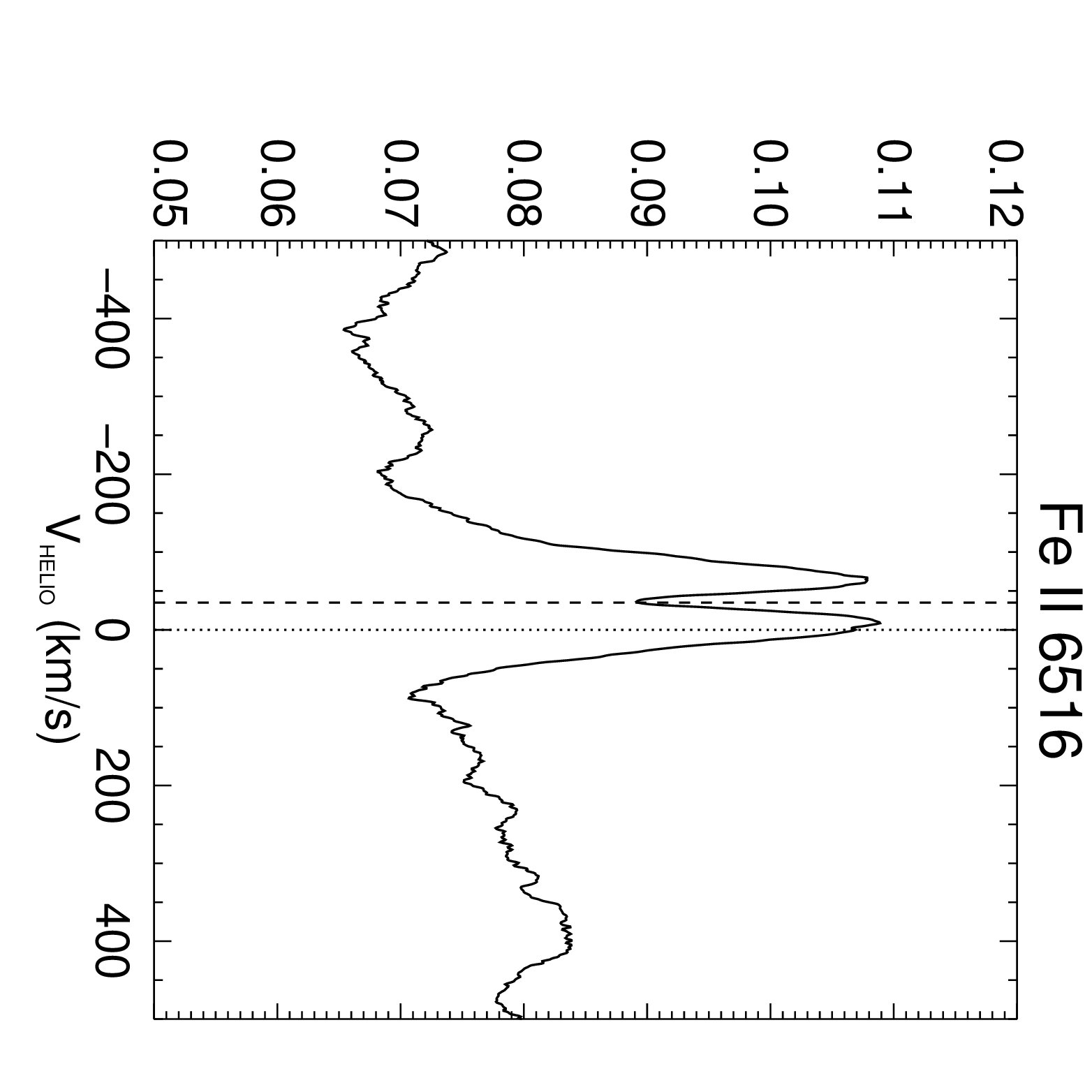}
\caption{The profiles of  H$\alpha$, H$\beta$, H$\gamma$, H$\delta$, HeI 5875, HeI 7065, FeII 4923 and FeII 6516 in the combined average spectrum. The spectra have not been corrected for the systemic velocity (-35 km s$^{-1}$, dashed line).}
\label{pic:pr}
\end{center}
\end{figure*}

The hydrogen Balmer lines and Fe II lines show a complex profile composed of superimposed emission and absorption components.
The absorption component is blue-shifted.
The traditional picture of symbiotic binaries is that the blue-shifted absorption is due to self-absorption by the circumbinary material due to the red giant stellar wind \citep{ikedatamura}.

We find that the Fe II lines can be divided into two classes: in one the centre of the absorption is blueshifted about -38 km s$^{-1}$ and in the other around -50 km s$^{-1}$.
While the H Balmer lines and Fe II lines have both emission and absorption components, the He I lines show only single peaked emission.
Figure \ref{pic:pr} shows the profiles of some lines of the combined spectrum of RS Oph.

No lines show double peaked emission profiles typical for accretion discs.
The apparent double peaked nature of the H and Fe II lines is due to an overlaid absorption feature, since the trough in the middle of the line profile extends well below the continuum level in several line profiles.
Apart from this, the H$\alpha$ line profiles resemble remarkably some of the polar inflow model profiles with low opening angle and intermediate accretion disc scattering optical depth \citep[Fig. 5,][]{booth}.
The absorption feature is thought to be due to the stellar wind of the red giant.
However, the inclination of RS Oph is probably too low for double peaked line profiles to appear despite the existence of an accretion disc.
The inclination of RS Oph has been estimated to be 39$^\circ$ by \citet{ribeiro}$, 49 - 51^\circ$ by \citet{brandi} and $\leq 35^\circ$ by \citet{dobrzycka94}.

Figures \ref{pic:hseq} -- \ref{pic:naid1seq} in the appendix show how the individual line profiles vary over the observations, and thus the orbital period.

\begin{figure}
\begin{center}
\includegraphics[width=0.45\textwidth]{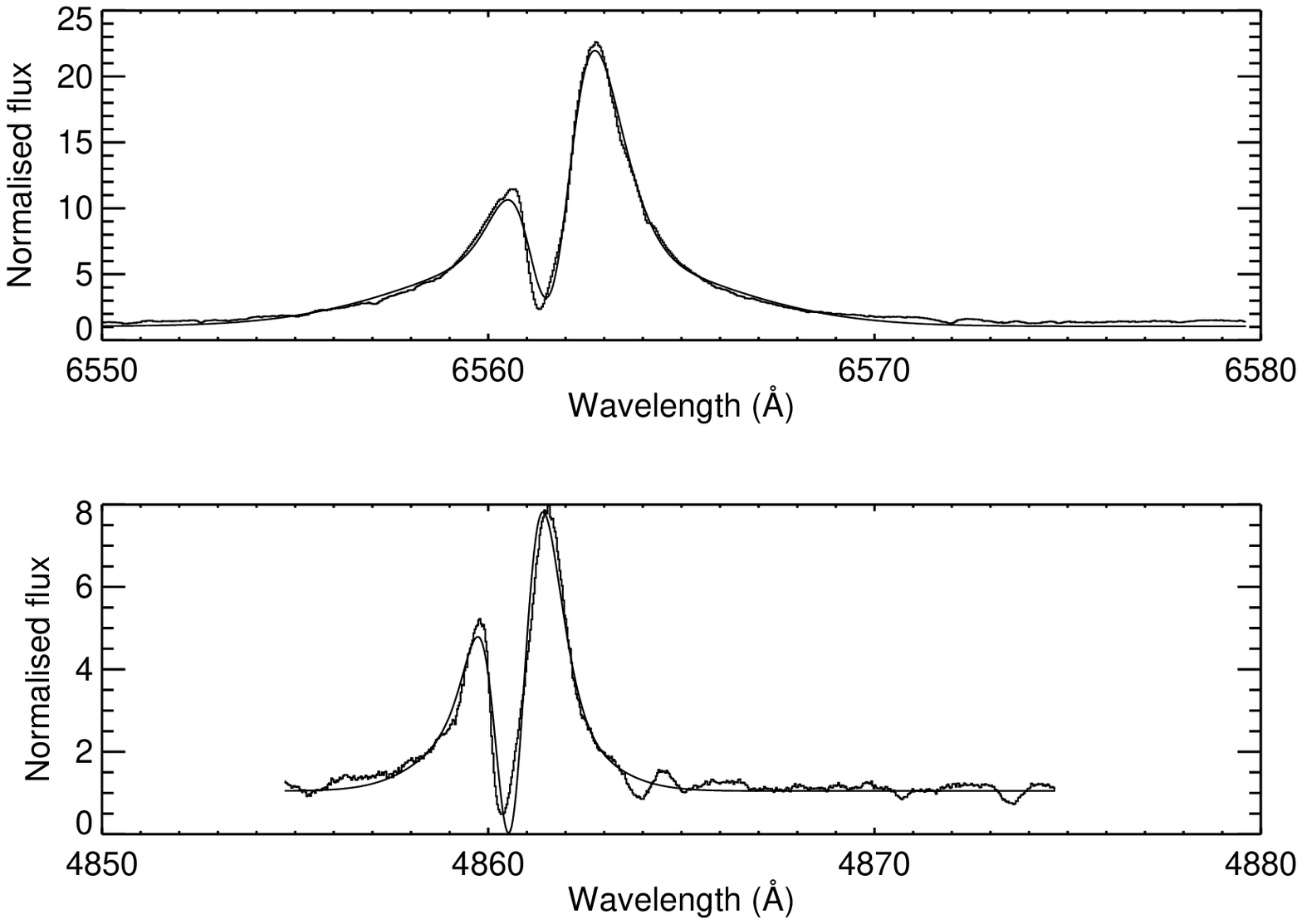}
\caption{Examples of H$\alpha$ (top) and H$\beta$ (bottom) profile fitting with three Gaussian components. Flux is relative to normalised continuum.}
\label{pic:proffit}
\end{center}
\end{figure}

To study the variability of H$\alpha$ and H$\beta$ the lines were fitted with a Gaussian multicomponent model: an absorption and two emission components: one broad and one narrow.
In order to model the variable emission lines, we fitted all the 30 line profiles of each line simultaneously.
This enables us to fit the line profiles with the same average absorption profile (determined during the fitting process), whilst still fitting the changing emission line profiles.
The motivation for this is that the absorption does not seem to change with the orbital phase and is thus likely to originate closer to the observer.
As a result the broad emission component did not show any specific periodical movement, but the narrow component showed sinusoidal variability.
Figure \ref{pic:proffit} shows examples of the fits of H$\alpha$ and H$\beta$ lines.

The sinusoidal variation of the narrow emission components was fitted with a radial velocity curve
\begin{equation}
\label{eq:radvel}
v_{rad} = \gamma + K \sin (2\pi \phi)
\end{equation}
resulting to a systemic velocity of about $\gamma = -35$ km s$^{-1}$ and semiamplitude of approximately $K = 6$ km s$^{-1}$ for both H$\alpha$ and H$\beta$ (see more detailed values in Table \ref{Tab:RadVel}).
The data with fits are shown in Figure \ref{pic:rvel}.
When correcting for the systemic velocity of the fit, the absorption component velocity becomes -18 km s$^{-1}$, which is a reasonable value for a stellar wind of a red giant \citep{espey, mccray}.

\subsection{Radial velocity curves}

In addition to the fits explained previously, we also determined the radial velocity curves of the components of RS Oph by two different methods.

\begin{figure*}
\begin{center}
\includegraphics[width=0.45\textwidth]{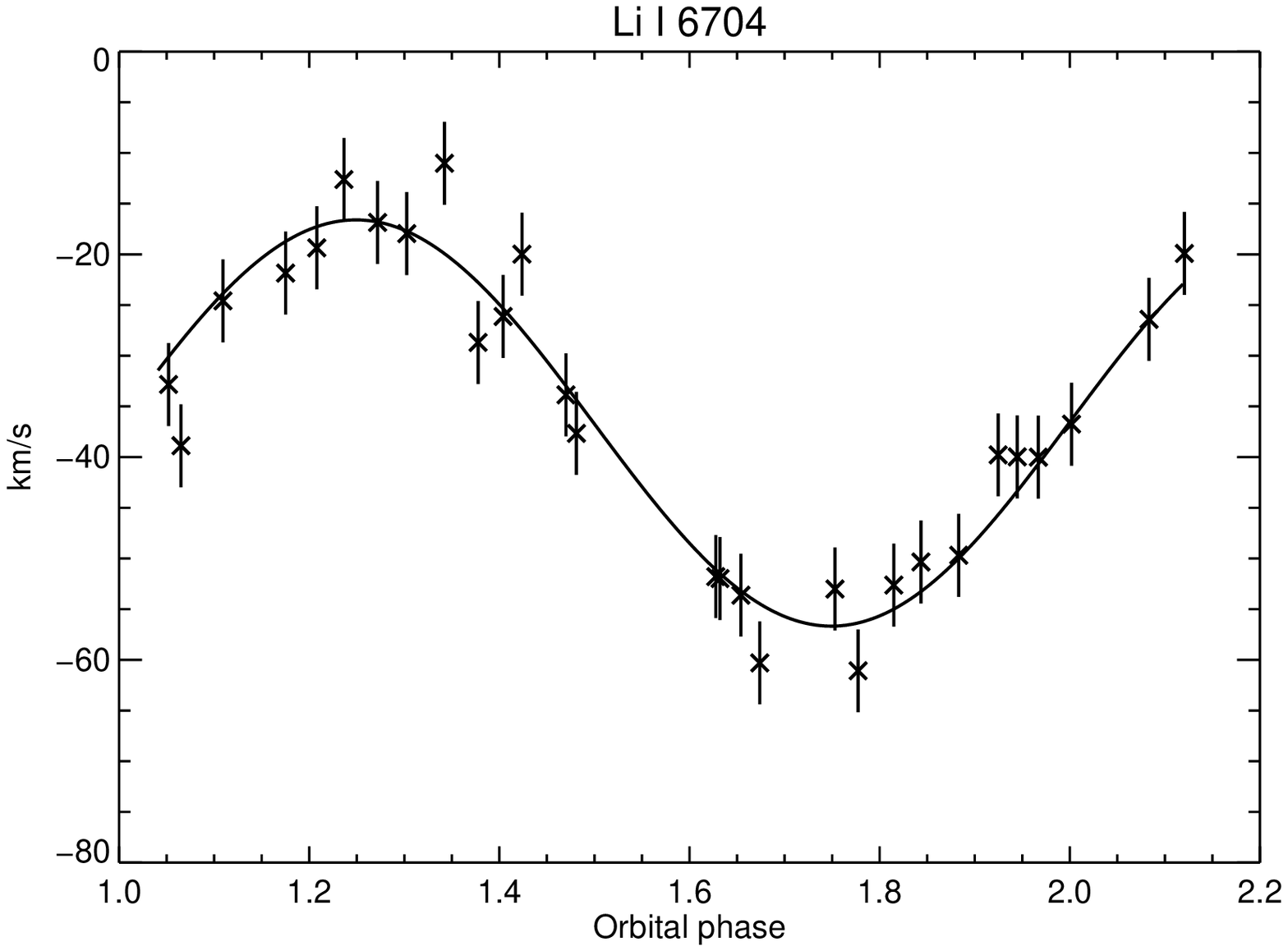}
\includegraphics[width=0.45\textwidth]{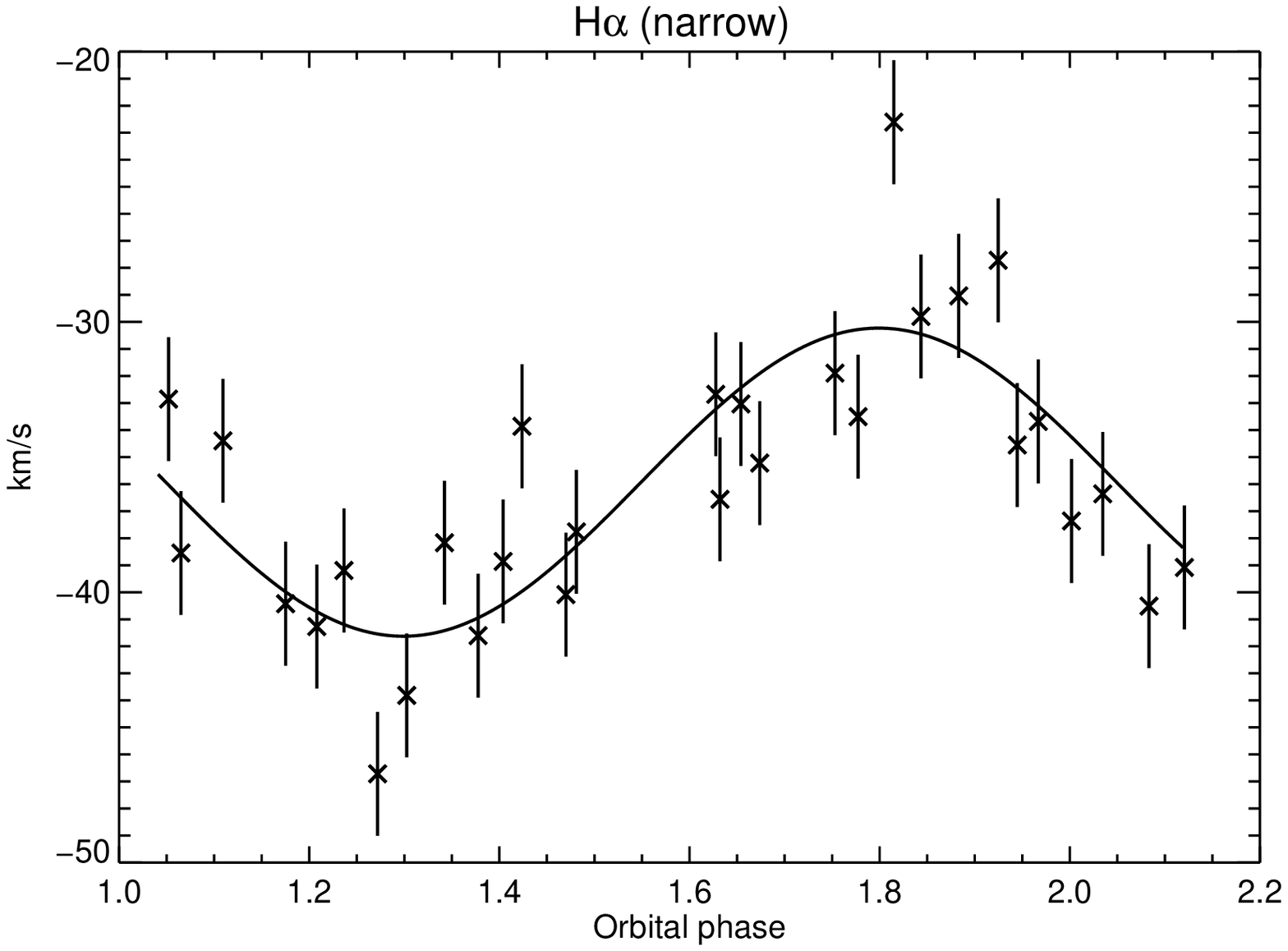}
\includegraphics[width=0.45\textwidth]{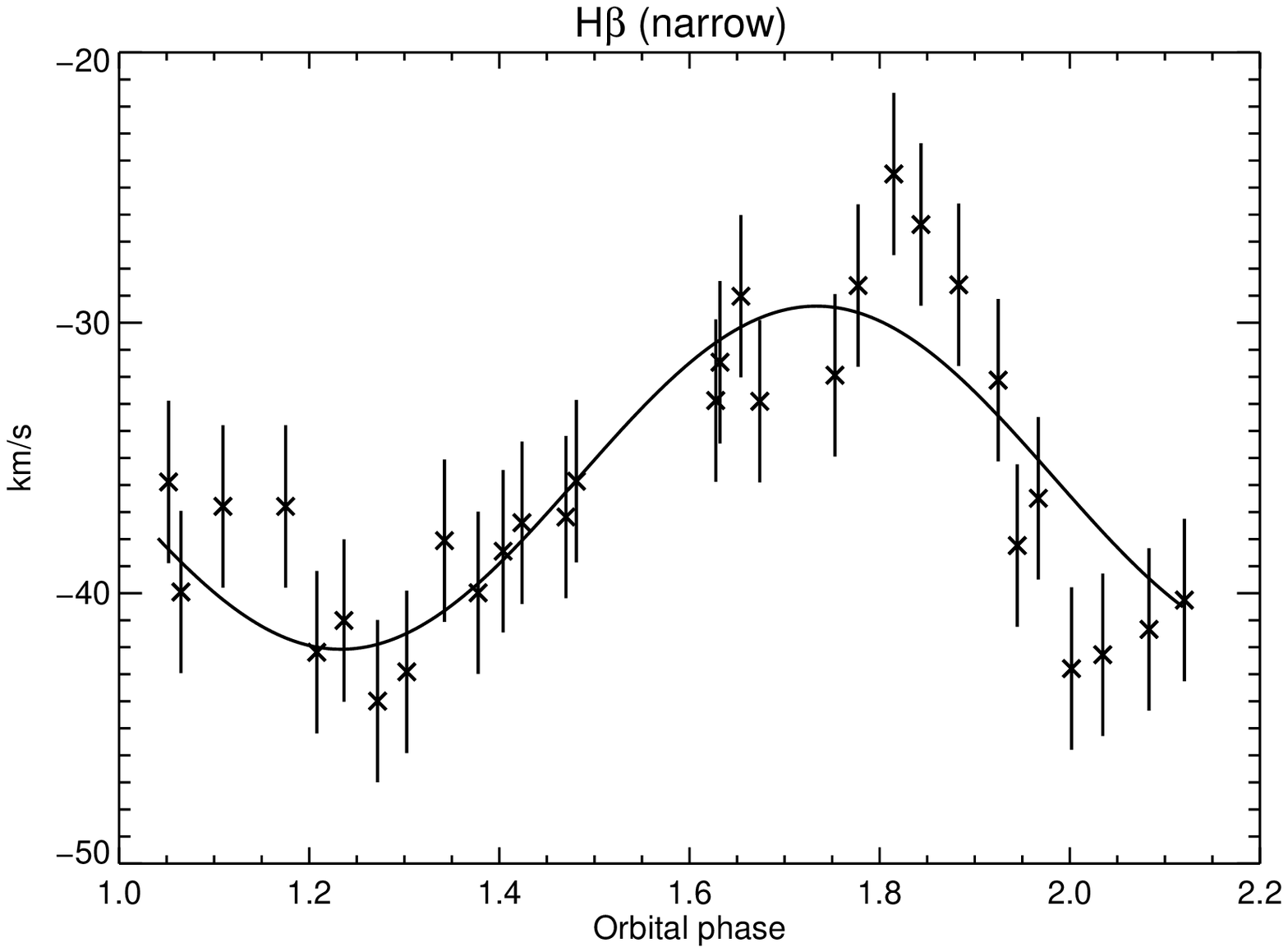}
\includegraphics[width=0.45\textwidth]{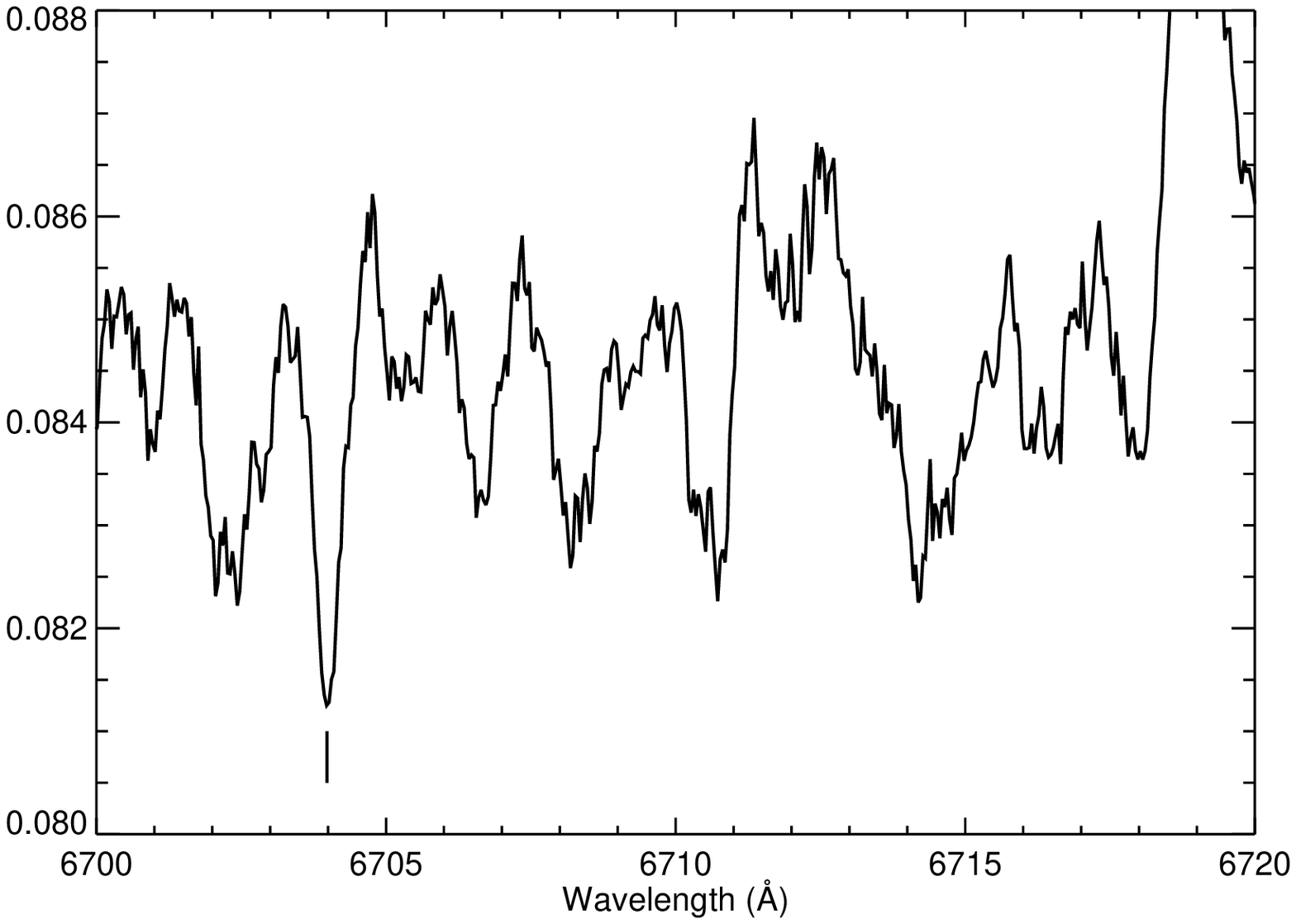}
\caption{Radial velocity curves. \textit{Top left}: Li I 6704\AA\ absorption feature, tracing the movement of the red giant. $\gamma = - 36.64\pm 0.52, K = 20.03 \pm 1.04 $ km s$^{-1}$ ($\chi^{2}_{\nu}$ = 1.12). The data point from 25 July 2009 has been omitted. 
\textit{Top right}: H$\alpha$ narrow emission component. $\gamma = -35.93 \pm 0.43, K = 5.71 \pm 0.85$ km s$^{-1}$
\textit{Bottom left}: H$\beta$ narrow emission component. $\gamma = -35.79 \pm 0.51, K = 6.58 \pm 1.00$ km s$^{-1}$.
\textit{Bottom right}: Average combined spectrum around the Li I line. Heliocentric radial velocities of each spectrum were reduced before combining. The mark shows the line at 6703.98 \AA\ that was fitted to measure the rotation of the red giant.
}
\label{pic:rvel}
\end{center}
\end{figure*}

The radial velocity curve of the red giant was calculated by cross correlating spectral regions with many absorption lines.
We decided on using a spectral region around the Li I 6707 line.
The spectra were cross correlated with the IRAF task FXCOR.
As we did not have a template spectrum, the cross correlations were done against one spectrum of a good signal-to-noise ratio (30 March 2009) and the resulting radial velocity curves were scaled to the correct systemic velocity using the systemic velocity of the H$\alpha$ narrow emission component fitting (which is in good agreement with the values of previous studies).
The radial velocity curves were again fitted with the function of Equation \ref{eq:radvel}.
One data point (25 July 2009) was omitted from the fit because the value deviated so much from all the others (a few times larger than the adjacent values with a ten times larger error).
This is probably due to bad signal to noise in the spectrum.
The results are shown in Fig. \ref{pic:rvel}.
The H$\alpha$ emission line wings are thought to be emitted close to the WD, so they should trace the radial velocities near the WD.
To determine the radial velocity curve for the H$\alpha$ wings we tried the method by \citet{schneider} implemented in the MOLLY software.
But this method is probably inappropriate for our data because it assumes the lines to be double peaked.

We also determined the radial velocity curves of the He I lines at 5875 \AA\ and 7065 \AA\ by fitting a Gaussian profile to the lines to see how the centre of the lines vary over the orbital period.
The FWHM of the fitted profile was 100 km s$^{-1}$ for both lines.
The orbital phases of the He I radial velocity curves are similar to the hydrogen Balmer lines.
Thus we suggest that the He I emission also originates in the vicinity of the WD.
The results of all radial velocity fits are shown in Table \ref{Tab:RadVel}.

\begin{figure}
\begin{center}
\includegraphics[width=0.4\textwidth, angle=270]{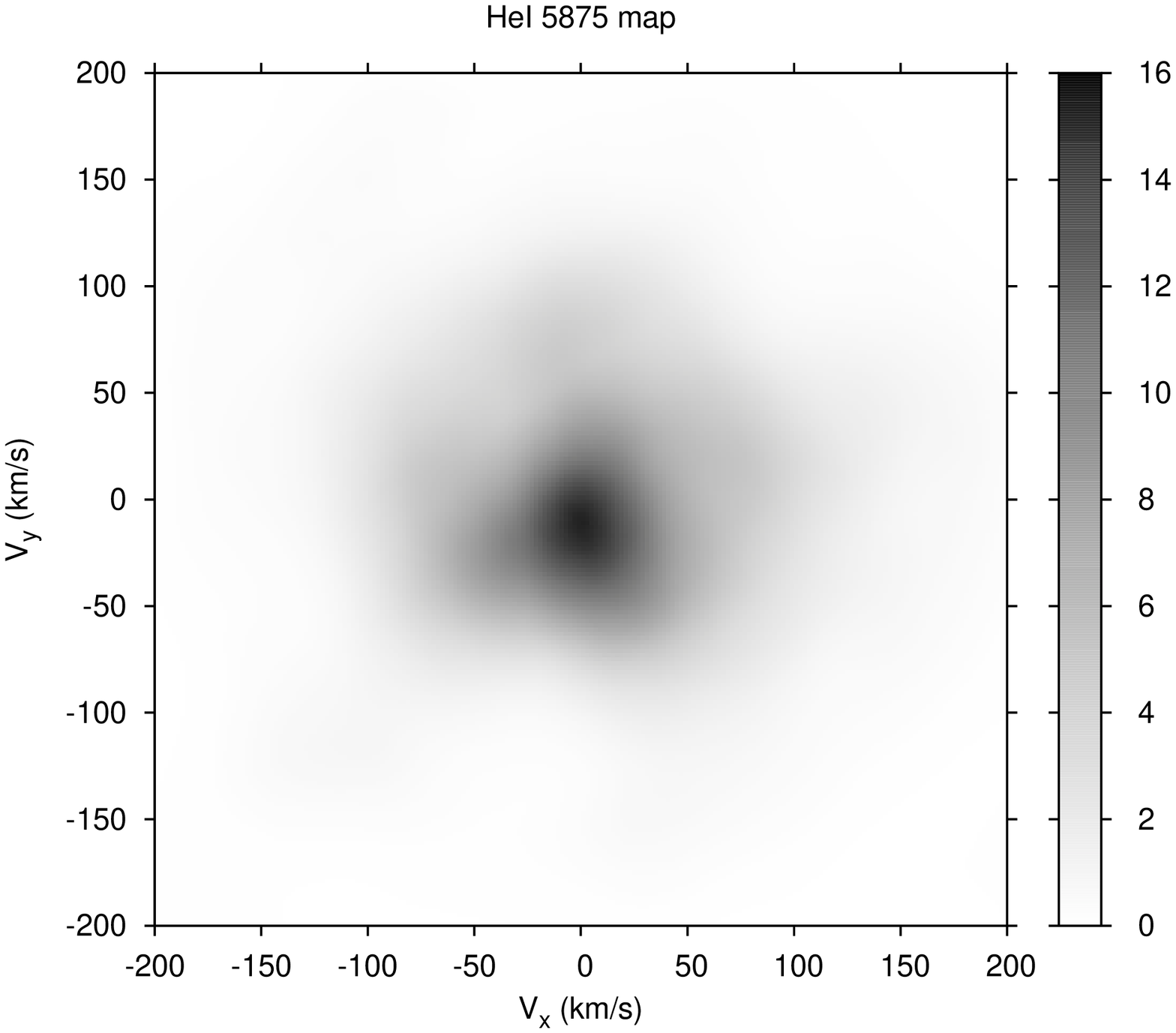}
\caption{Doppler map of He I 5875 line.}
\label{pic:HeImap}
\end{center}
\end{figure}

We also applied Doppler tomography \citep{marsh} to the He I emission lines and obtained a map of the emission distribution in velocity space.
Due to the presence of the absorption component, the technique could not be applied to the hydrogen lines.
However, we acknowledge that one should be cautious about interpreting the map as the technique assumes that all motion is restriced in the orbital plane which is probably not the case in RS Oph.
The map of He I 5875 is shown in Fig. \ref{pic:HeImap}.
It does not show any ring-like structure which could be interpreted as an accretion disc.
This is an expected result, as the line profiles are not double peaked, but the emission seems to be concentrated in a region around the white dwarf.

\begin{table}
 \centering
 \begin{minipage}{140mm}
  \caption{Measured radial velocities.}
  \label{Tab:RadVel}
  \begin{tabular}{@{}lll@{}}
  \hline
 & $\gamma$ (km s$^{-1}$) & $K$ (km s$^{-1}$)\\
 \hline
H $\alpha$ (narrow component) & $-35.93 \pm 0.43$ & $5.71 \pm 0.85$\\
H $\beta$ (narrow component) & $-35.79 \pm 0.51$ & $6.58 \pm 1.00$\\
H$\alpha$ wings & $-33.15 \pm 3.85 $& $29.61 \pm 5.53$\\
He I 5875 center & $-21.39 \pm 1.11$& $8.17 \pm 1.50$ \\
He I 7065 center & $-28.37 \pm 0.85$ & $5.67 \pm 1.16$\\
Li I & $-36.64\pm 0.52$ & $20.03 \pm 1.04$\\
\hline
\end{tabular}
\end{minipage}
\end{table}

\subsection{Mass ratio}

We can calculate the mass ratio of RS Oph from the semiamplitudes of the radial velocities.
The radial velocity of Li I is used as a tracer of the red giant movement.
For the WD we use the narrow emission of the H$\alpha$ line as we assume the narrow component to originate close to the WD.
This gives a mass ratio $q=M_{RG}/M_{WD}=0.285 \pm 0.045$.
When using the He I 5875 line as a tracer for the WD, the mass ratio becomes $q=0.408 \pm 0.078$.
If we assume the mass of the WD to be $1.2 - 1.4$ M$_{\odot}$ we get a mass of the red giant to be $0.342 - 0.399$ M$_{\odot}$ and $0.489 - 0.571$ M$_{\odot}$ with the lower and higher mass ratio, respectively.
Our mass ratio is smaller than the value previously reported \citep[$q=0.59\pm 0.05$,][]{brandi}.
In the following we adopt a value of $q=0.285$.
We choose this value because the error is smaller in the H$\alpha$ fit than in the He I fit.

\subsection{Rotation of the red giant}

To study the rotation of the red giant, the heliocentric velocities as well as the orbital radial velocities (based on the fit of the radial velocity curve of the red giant, see Fig. \ref{pic:rvel}) were removed from the spectra and then they were combined.
The FWHM width of the absorption feature of Fe I at the wavelength 6703.98 \AA\ in the combined spectrum was measured to be 0.36 \AA.   
After correcting for the instrumental boardening of 6.2 km s$^{-1}$, measured from an arc line close to the absorption feature, we obtain a projected velocity of $v \sin i =$ 14.8 km s$^{-1}$.
The line width cannot be caused by atmospheric turbulence only.
\citet{pavlenko2010} determined the microturbulent velocity of the RG to be 3 km s$^{-1}$.
The convection induced macroturbulence in late type stars is also only of order of a few km s$^{-1}$ \citep{gray}.
As the measured line width is an order of magnitude higher, we interpret it being caused by the rotation of the red giant.

\begin{figure}
\begin{center}
\includegraphics[width=0.5\textwidth, angle=0]{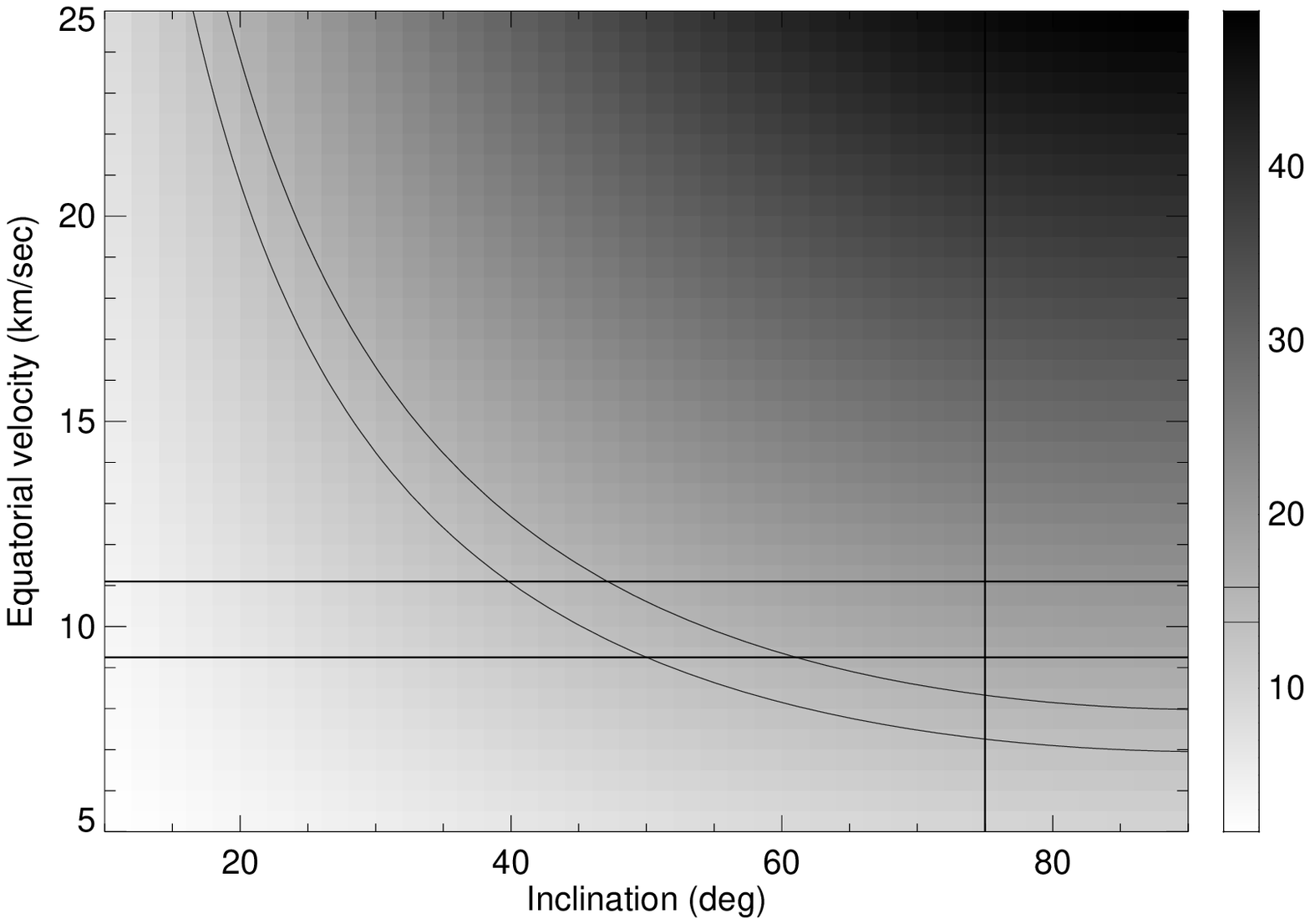}
\caption{Result of simulating possible observable FWHM values. Velocity at the equator of the red giant against the orbital inclination angle. Gray scale is the value of FWHM. As the system does not show any eclipses, the inclination angle has to  be smaller than 75$^{\circ}$ (marked with vertical line). The black contour lines limit the region where the observed FWHM is between 13.8 and 15.8 km s$^{-1}$. The higher horizontal line corresponds to the equatorial velocity of calculated with a mass ratio of $q=0.59$.}
\label{pic:spin}
\end{center}
\end{figure}

Figure \ref{pic:spin} shows the dependence of observed FWHM as a function of equatorial velocity and the system inclination.
As no eclipses are observed in RS Oph, the inclination must be less than $\sim75^{\circ}$ for a mass ratio of $q = 0.285$ (marked with the dotted line).
The two black solid lines in the plot limit the region where the observed FWHM is between 13.8 and 15.8 km s$^{-1}$ (as our measurement was 14.8 km s$^{-1}$).
So the possible values for the equatorial velocity are in the region limited by the lines and below inclination of 75$^{\circ}$.

We calculate the Roche lobe radius of the red giant with the equation by \citet{eggleton}

\begin{equation}
\label{eq:rochel}
\frac{R_2}{a} = \frac{0.49 q^{2/3}}{0.6 q^{2/3} + \ln(1+q^{1/3})}
\end{equation}
where $R_2$ is the Roche lobe radius, $q$ the mass ratio and $a$ the binary separation, and (using $q = 0.285$) get the value $ R_2 = 5.77 \times 10^{10}$ m.
If we assume that the red giant fills its Roche lobe, take into account that the orbital period is 456.3 d, and assume a synchronous rotation, we get an equatorial velocity of 9.25 km s$^{-1}$ (ignoring all other effects).
This velocity is marked with a line in Fig. \ref{pic:spin}.

As we can see, this value fits well with the observed FWHM if the inclination is between 50$^{\circ}$ and 60$^{\circ}$.
Thus we suggest that the red giant is tidally locked in the system.

\subsection{Sodium lines and link to Ia supernovae}

RS Oph is a strong candidate for a SN Ia progenitor \citep{hernanz, starrfield}.
In a recurrent nova outburst the ejecta sweeps through the circumbinary matter (CBM)  pushing the matter originated from the red giant stellar wind away changing the density structure in the surrounding medium.
Signatures of this kind of pre-explosion density structure have been observed in some SNe Ia \citep[e.g.][]{woodsokoloski}.

\citet{iijima} and \citet{patat} have detected components of the Na I D1 line in RS Oph spectrum that cannot be explained by interstellar absorption but are suggested to originate in the CBM.

\citet{patat} distinguished 5 different components in the Na I D1 line: at velocities of -77 (\#1), -63 (\#2), -46 (\#3), -12 (\#4) and +2 (\#5) km s$^{-1}$.
They use a systemic velocity of -40.2 km s$^{-1}$ and suggest that components \#1 -- \#3 arise in the CBM.
Their data consists of only a few epochs before and after the 2006 outburst of RS Oph, whereas our data covers one whole orbital period.

The Na I line can be seen in Fig. \ref{pic:naid1seq}.
We can see that that the line profiles do not change depending on the orbital phase supporting the idea that blue-shifted components are indeed due to CBM that has been structured by previous nova eruptions.

\section{Polarisation spectrum}

We fitted the RS Oph polarisation spectrum with an empirical "Serkowski law" \citep{serkowski, serkowski2} which describes the interstellar polarisation as a function of wavelength:

\begin{equation}
\label{eq:serkowski}
\frac{P(\lambda)}{P(\lambda_{max})} = e^{-K\ln^2 (\lambda_{max}/\lambda)}
\end{equation}
where $P(\lambda_{max})$ is the maximum degree of polarisation, $\lambda_{max}$ is the wavelength where it occurs and $K$ is a constant determining the width of the function.

\begin{figure}
\begin{center}
\includegraphics[width=0.5\textwidth, angle=0]{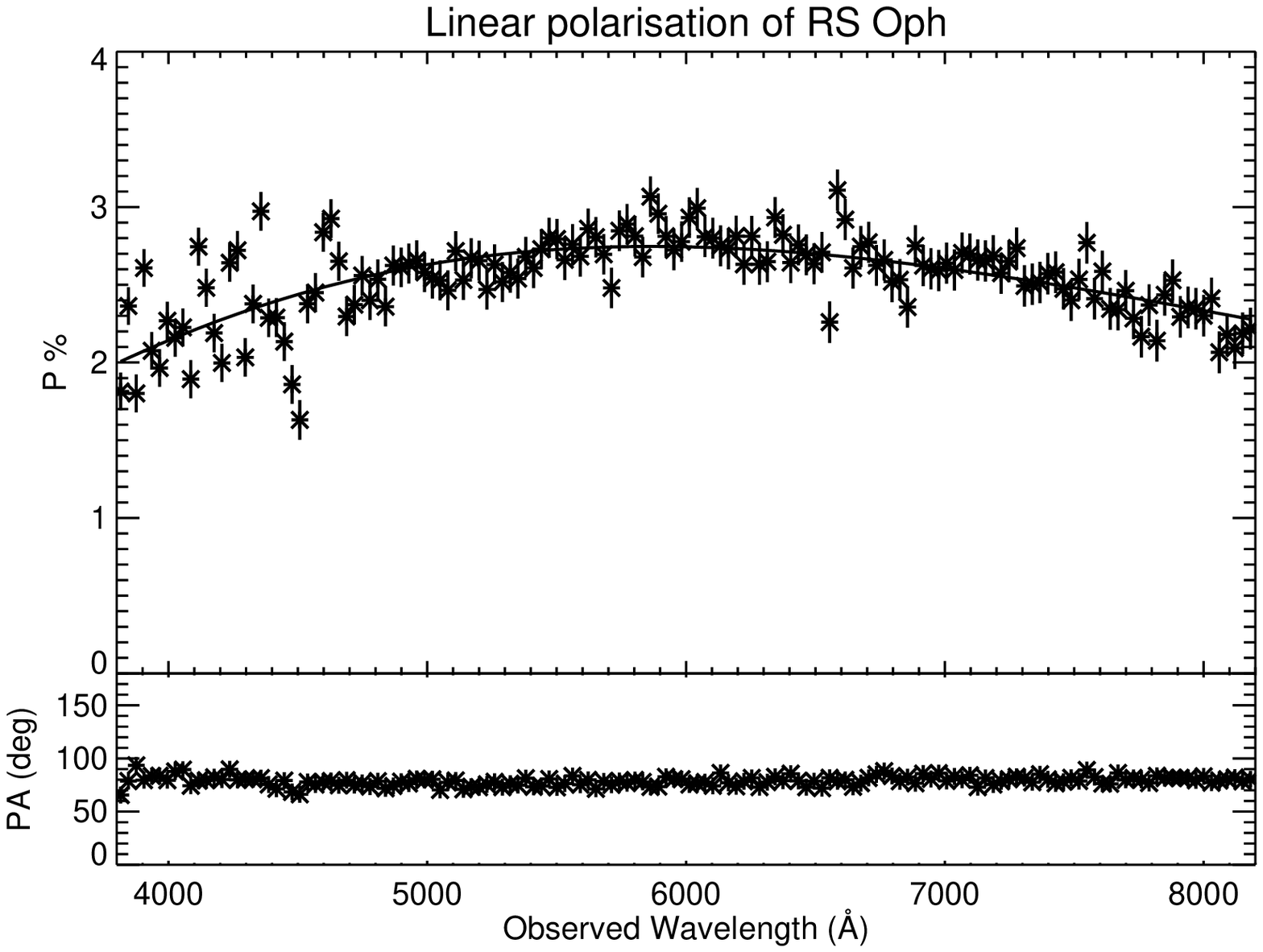}
\caption{Polarisation spectrum of RS Oph (+) and the Serkowski law fit. Upper panel shows the binned spectrum, lower panel shows the position angle.}
\label{pic:polspec}
\end{center}
\end{figure}

The result of the fit is presented in Table \ref{Tab:serkowski} and the binned polarisation spectrum together with the fit are shown in Figure \ref{pic:polspec}.
The average of the position angle is 78.7$^{\circ}\pm 5.1^{\circ}$.
As the Serkowski law fits quite well the polarisation spectrum, it seems that we observe only interstellar polarisation and we cannot detect any intrinsic polarisation.

Our result is in agreement with earlier observations by \citet{cropper} where they detected intrinsic polarisation only during the outburst and only interstellar polarisation during quiescence.
Also their values of the fit are similar to ours, apart from the constant K which we allowed as a free parameters while they used a fixed value of 1.15.

\begin{table}
 \centering
 \begin{minipage}{75mm}
  \caption{The parameters of the Serkowski law fit.}
  \label{Tab:serkowski}
  \begin{tabular}{@{}lc@{}}
  \hline
Parameter & Value\\
\hline
$P(\lambda_{max})$(\%) & 2.75 $\pm$ 0.02\\
$K$ & 1.69 $\pm$ 0.10\\
$\lambda_{max}$ (\AA) & 5867.94 $\pm$ 37.41\\
\hline
\end{tabular}
\end{minipage}
\end{table}

\section{Discussion}
We do not detect He II nor O III emission lines in our spectra.
The ionisation energy for both species is around 54 eV.
The absence of both lines hints that the temperature of the WD is lower during our observations than it was in 2004, before the last outburst, when \citet{zamanov} observed He II 4686.
However, they observed variations on a timescale of a day, which we would miss in our data.
Or the line is simply too weak and embedded in the continuum.

Even if we do not see direct evidence of an accretion disc, it is possible that an accretion disc exists and we just cannot detect disc-like features because of a low inclination angle that has been suggested for RS Oph.

We find the absence of an absorption component in the helium emission lines perplexing, as it is present in both hydrogen and iron.
Either the circumbinary matter does not absorb emission at the wavelengths of He I emission or, more likely, the emission of H and He originate in different locations.
In the latter case both hydrogen and iron emission would originate much ''deeper'' in the circumbinary matter, probably close to the WD, while helium emission would arise from the outer shell of this matter.

There seem to be two classes of Fe II lines: one that has an absorption component at the systemic velocity. 
And another where the absorption is blueshifted with respect to the systemic velocity.
The one at systemic velocity arises in the red giant atmosphere. 
The one with the blueshifted component is a combination of the RG component and absorption from the circumbinary matter.

The mass ratio we measure is lower than previously reported.

The Na I D1 line show features that can be interpreted as absorption due to circumbinary material.

\section{Conclusions}
Based on our data, we cannot confirm the existence of an accretion disc in RS Oph as we do not detect any supporting spectral features.
The mass transfer mechanism remains unclear.
Our data clearly points out that the hydrogen lines are not double peaked emission lines (as referred by authors of some previous articles) but a combination of emission and absorption.
In a double peaked line setting the middle dip would not go under the continuum level like in our data in all hydrogen Balmer series lines.

We conclude that, if the width of the absorption line Fe I 6703.98 \AA\ can be interpreted being due to rotation of the red giant, the rotation of the red giant is consistent with its being tidally locked.

We do not detect any intrinsic polarisation from RS Oph, the linear polarisation we observe is explained by interstellar matter in the line of sight to the target.

\section*{Acknowledgements}
A.S. acknowledges funding from the European Commission under the Marie Curie Host Fellowships Action for Early Stage Research Training SPARTAN programme (Centre of Excellence for Space, Planetary and Astrophysics Research Training and Networking) Contract No MEST-CT-2004-007512, University of Leicester, UK; the Finnish Graduate School in Astronomy and Space Physics; and Academy of Finland grant 277375.
Based on observations made with the Nordic Optical Telescope, operated on the island of La Palma jointly by Denmark, Finland, Iceland, Norway, and Sweden, in the Spanish Observatorio del Roque de los Muchachos of the Instituto de Astrofisica de Canarias.  
The data presented here were obtained in part with ALFOSC, which is provided by the Instituto de Astrofisica de Andalucia (IAA) under a joint agreement with the University of Copenhagen and NOTSA.
We thank Tom Marsh for the use of MOLLY and DOPPLER software.
We are grateful to Dr. Sarah Casewell for useful comments on the manuscript of this paper.
We thank the anonymous referee for useful comments.




\bibliographystyle{mnras}
\bibliography{rsoph} 




\appendix

\section{Variability of emission lines}

\begin{figure*}
\begin{center}
\includegraphics[width=0.45\textwidth]{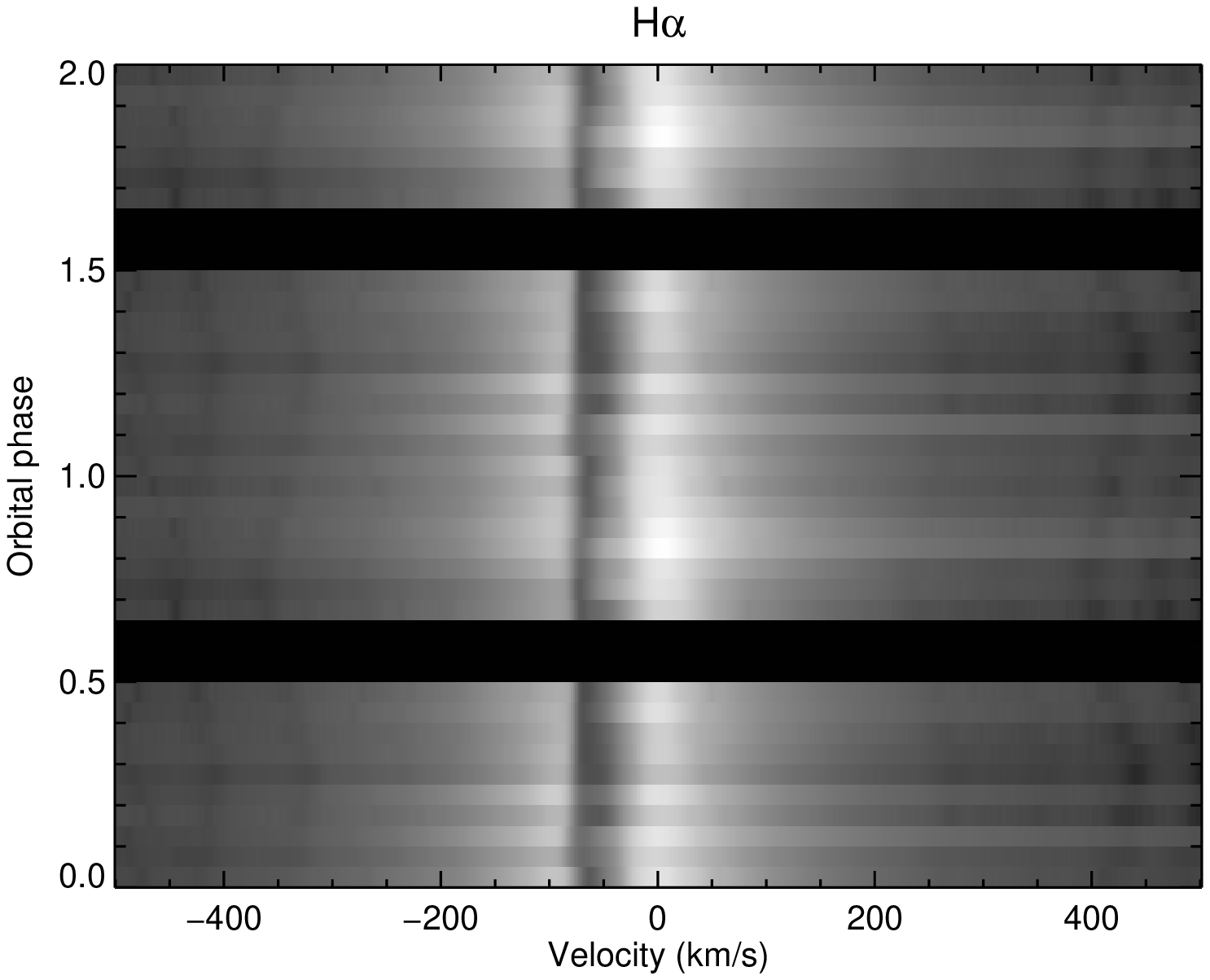}
\includegraphics[width=0.45\textwidth]{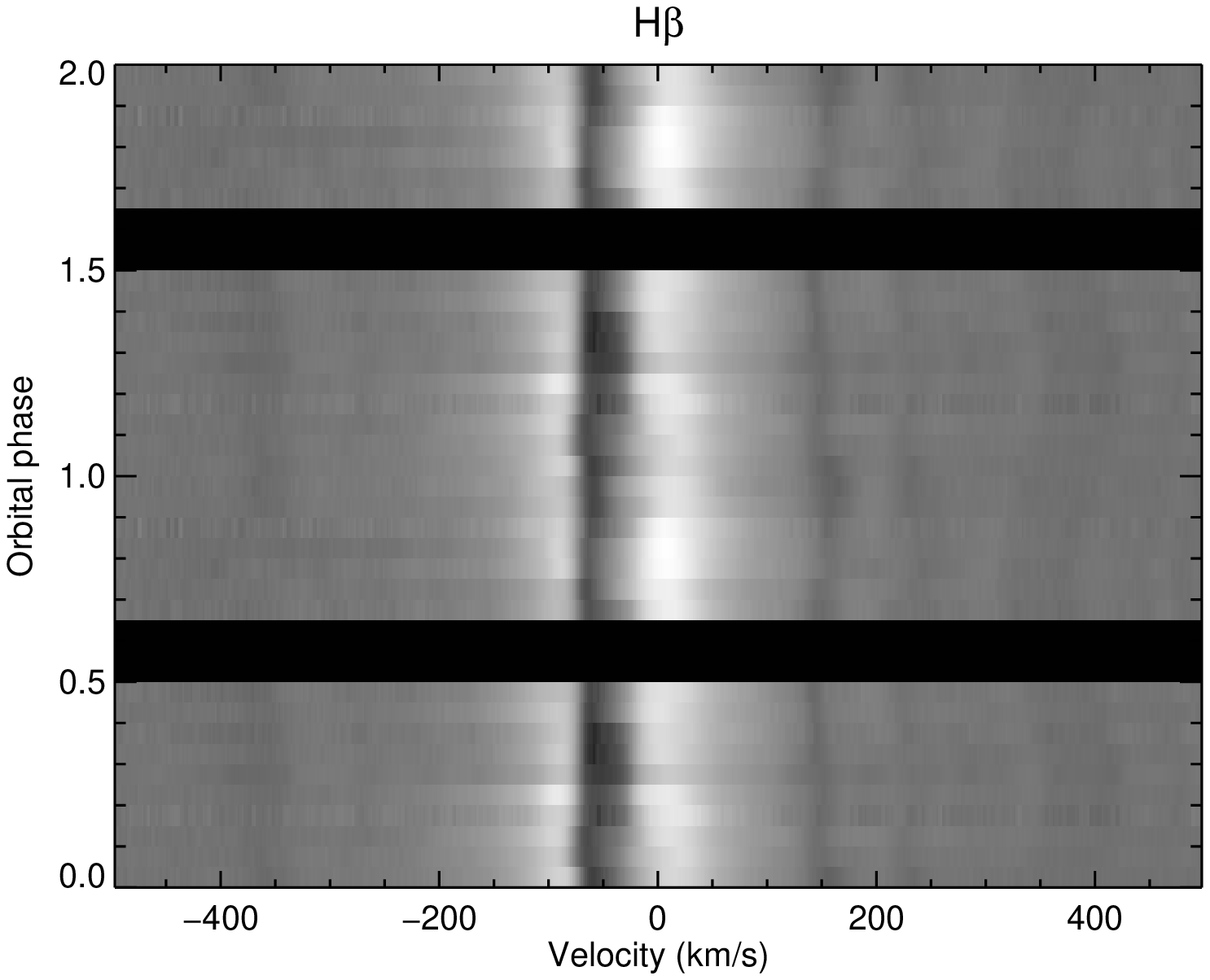}
\includegraphics[width=0.45\textwidth]{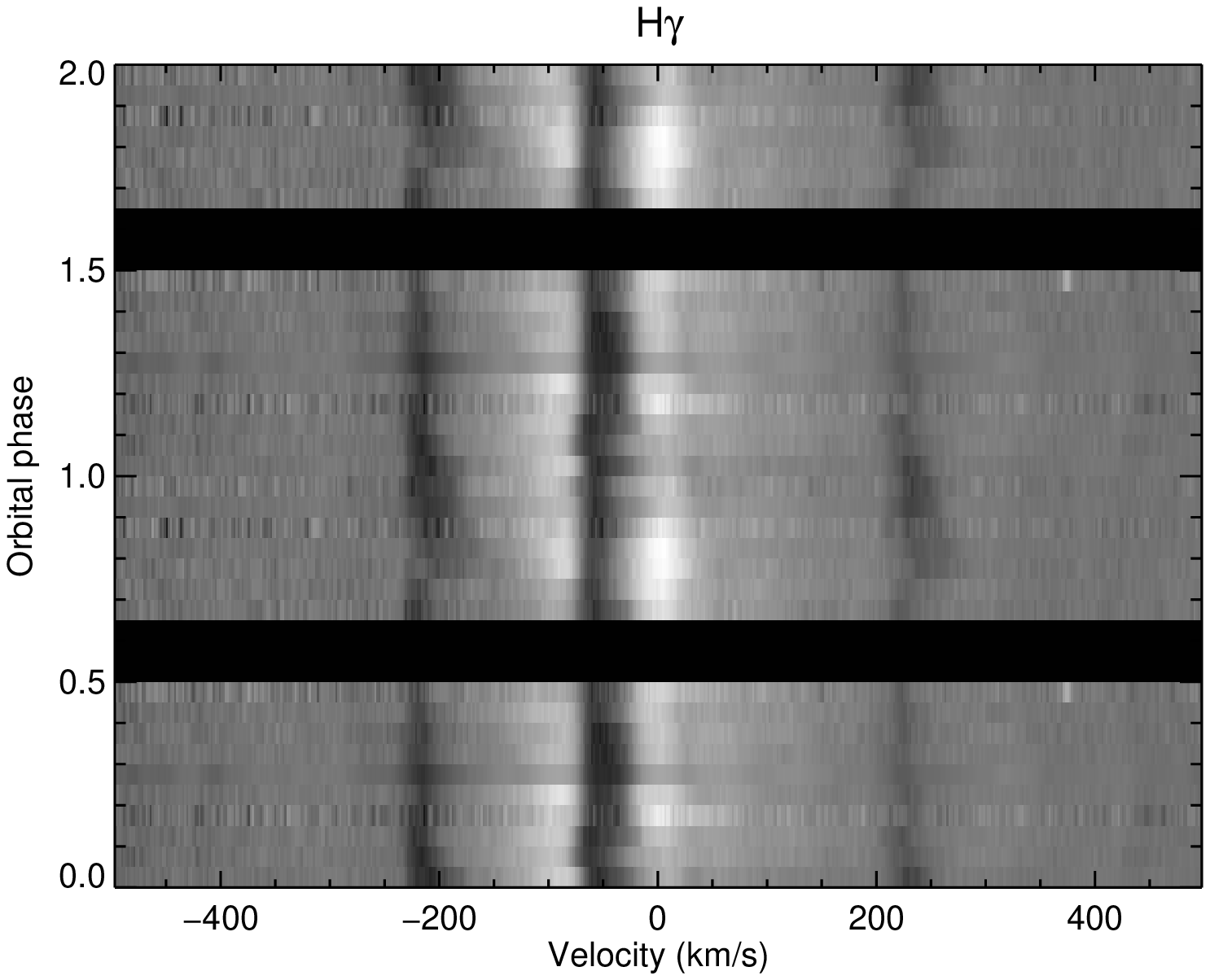}
\includegraphics[width=0.45\textwidth]{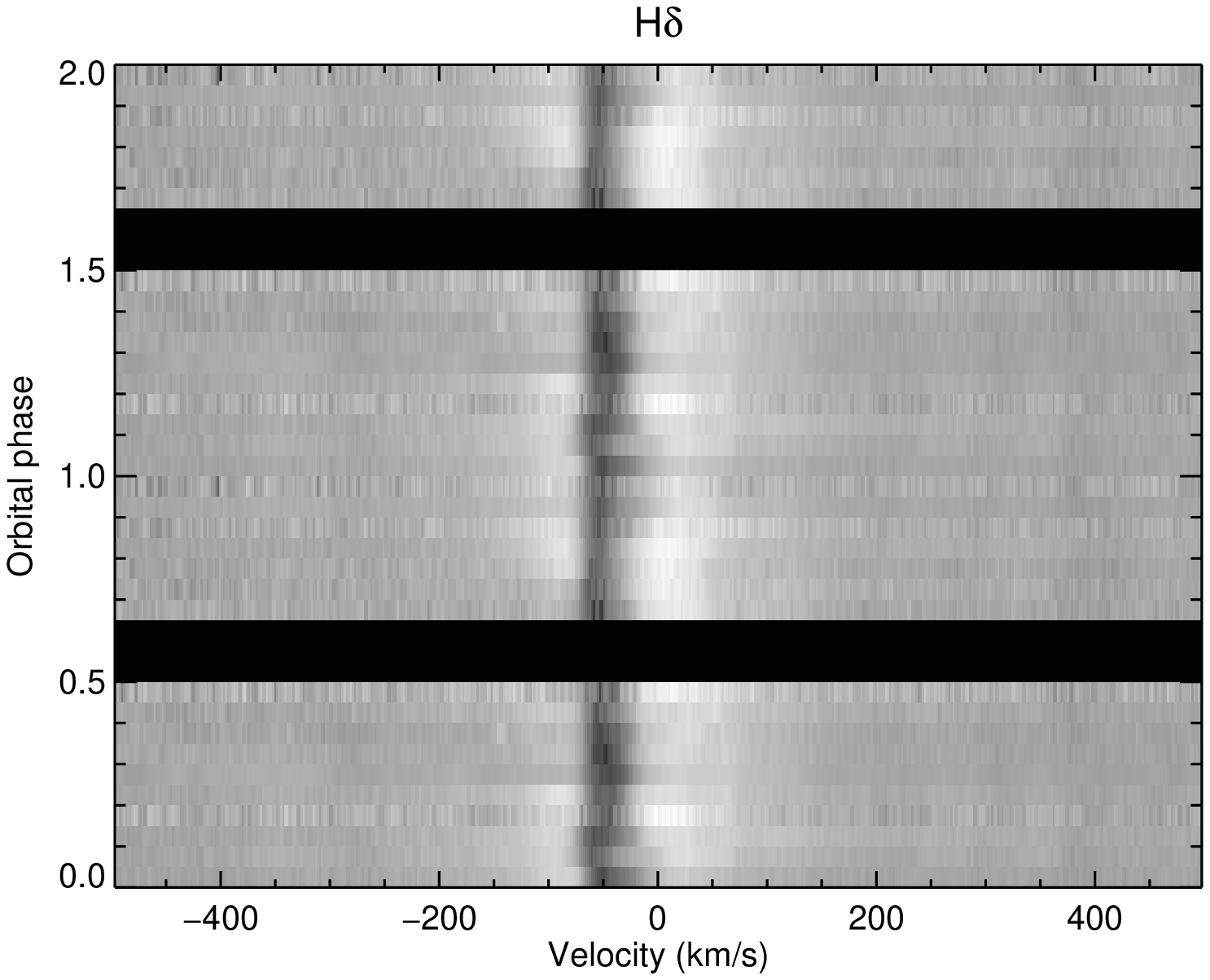}
\caption{The variability of H$\alpha$, H$\beta$, H$\gamma$ and H$\delta$ lines. Spectra have been normalised to continuum and phase binned into 20 bins and plotted over two orbital phases for clarity. The data has not been corrected for the systemic velocity (-35 km s$^{-1}$). Some faint absorption features from the red giant cause the s-like curves.}
\label{pic:hseq}
\end{center}
\end{figure*}

\begin{figure*}
\begin{center}
 \includegraphics[width=0.45\textwidth]{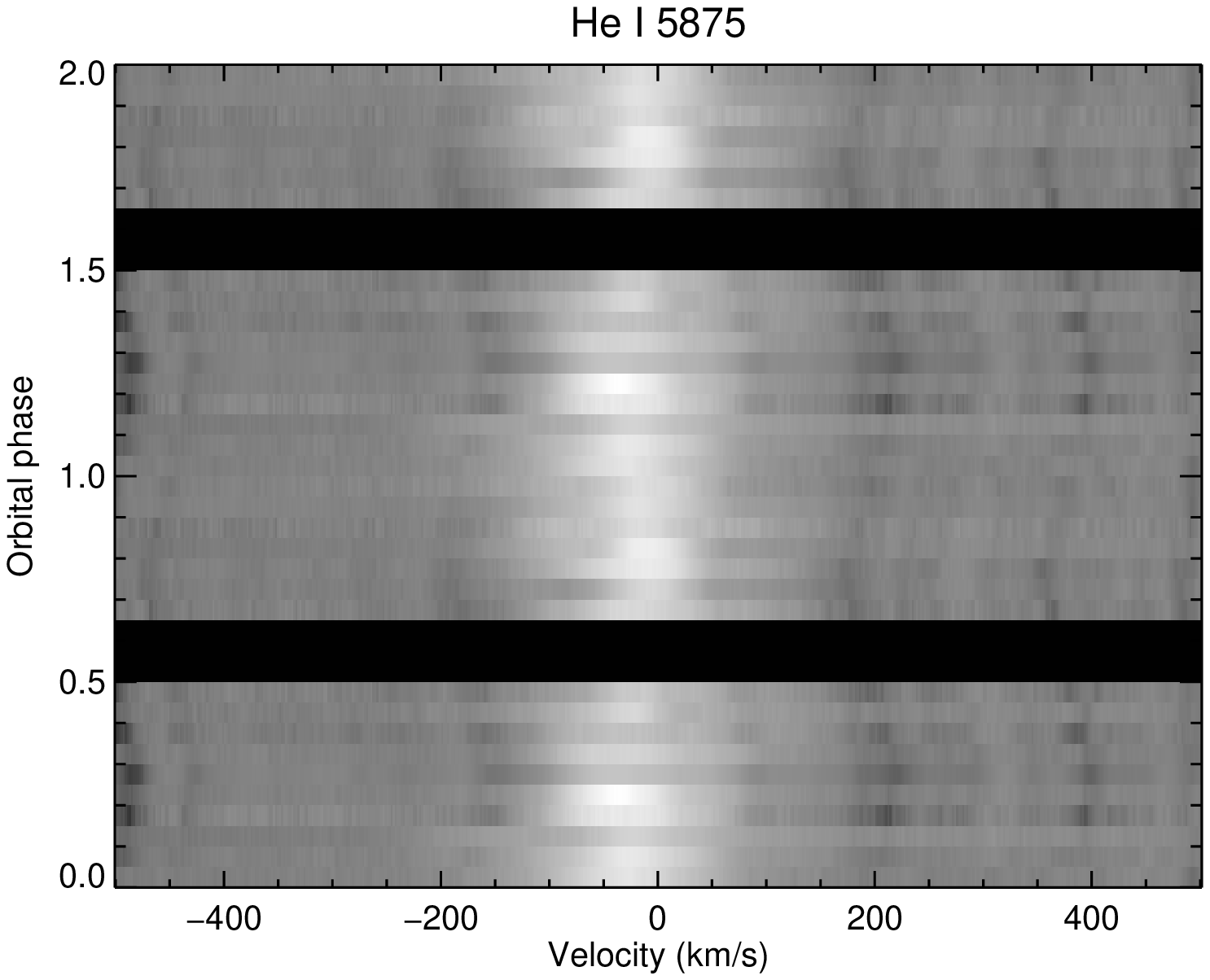}
 \includegraphics[width=0.45\textwidth]{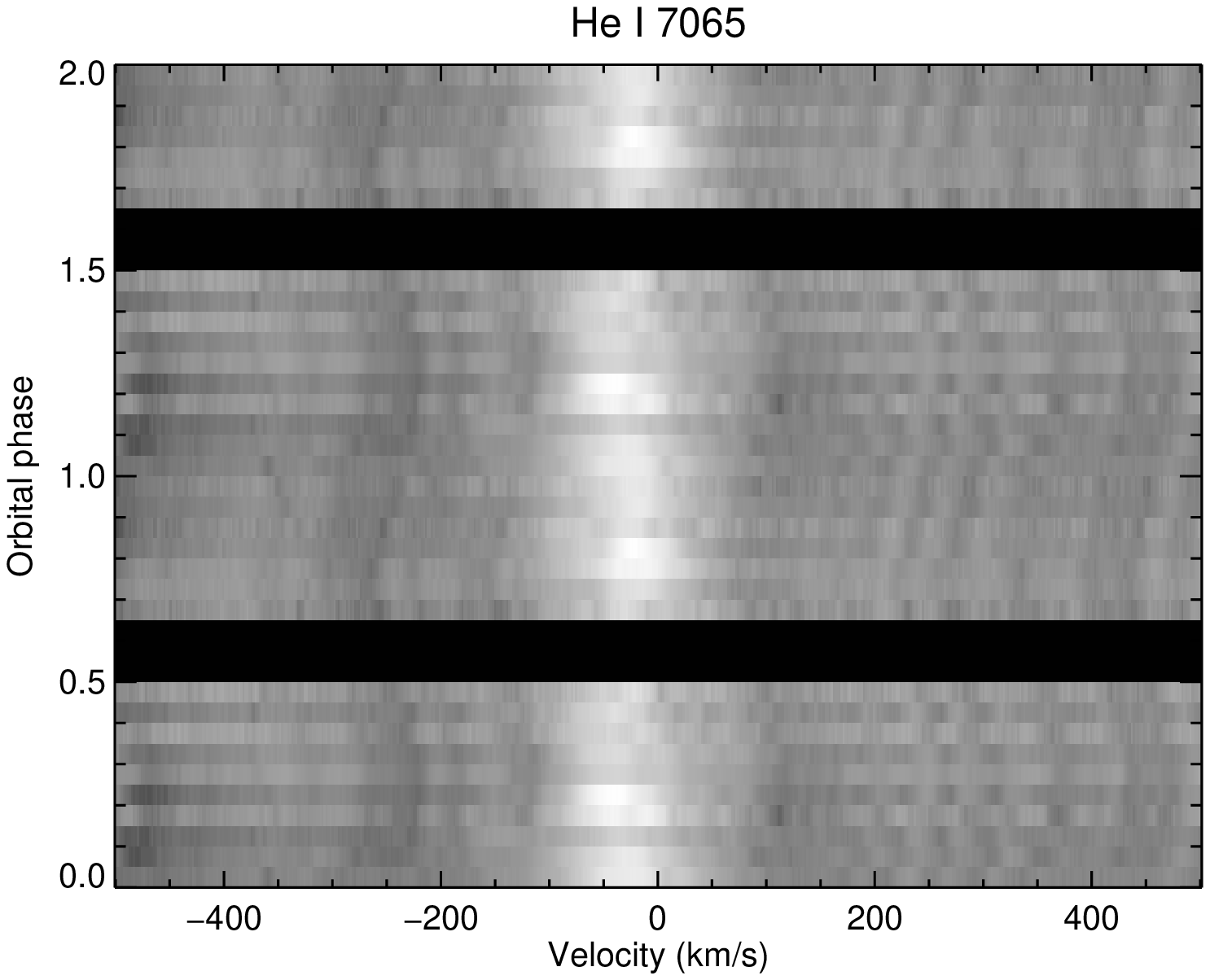}
\caption{The variability of HeI 5875 and HeI 7065 lines. Spectra have been normalised to continuum and phase binned into 20 bins and plotted over two orbital phases for clarity. The data has not been corrected for the systemic velocity (-35 km s$^{-1}$). Some faint absorption features from the red giant cause the s-like curves.}
\label{pic:heiseq}
\end{center}
\end{figure*}

\begin{figure*}
\begin{center}
 \includegraphics[width=0.45\textwidth]{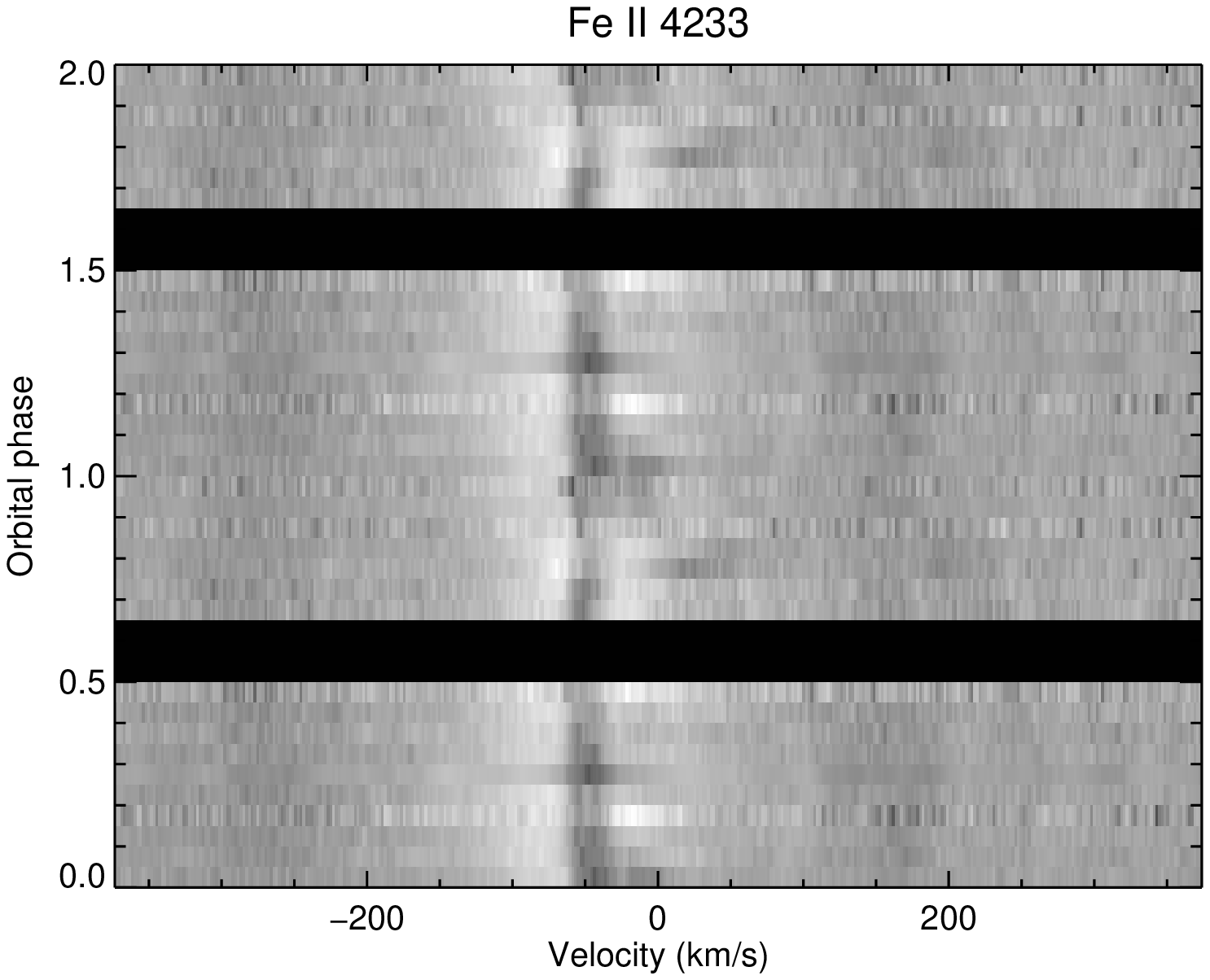}
 \includegraphics[width=0.45\textwidth]{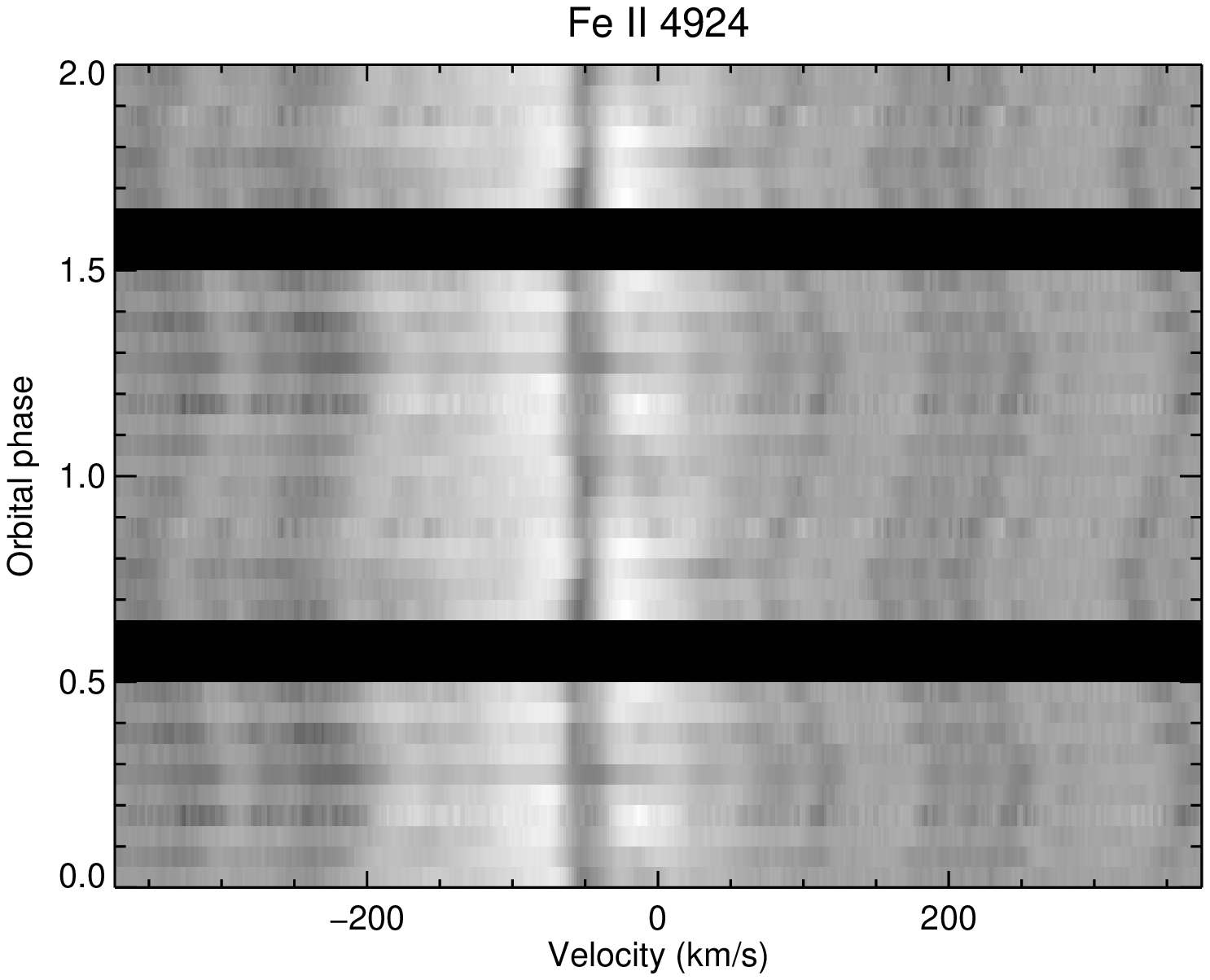}
\includegraphics[width=0.45\textwidth]{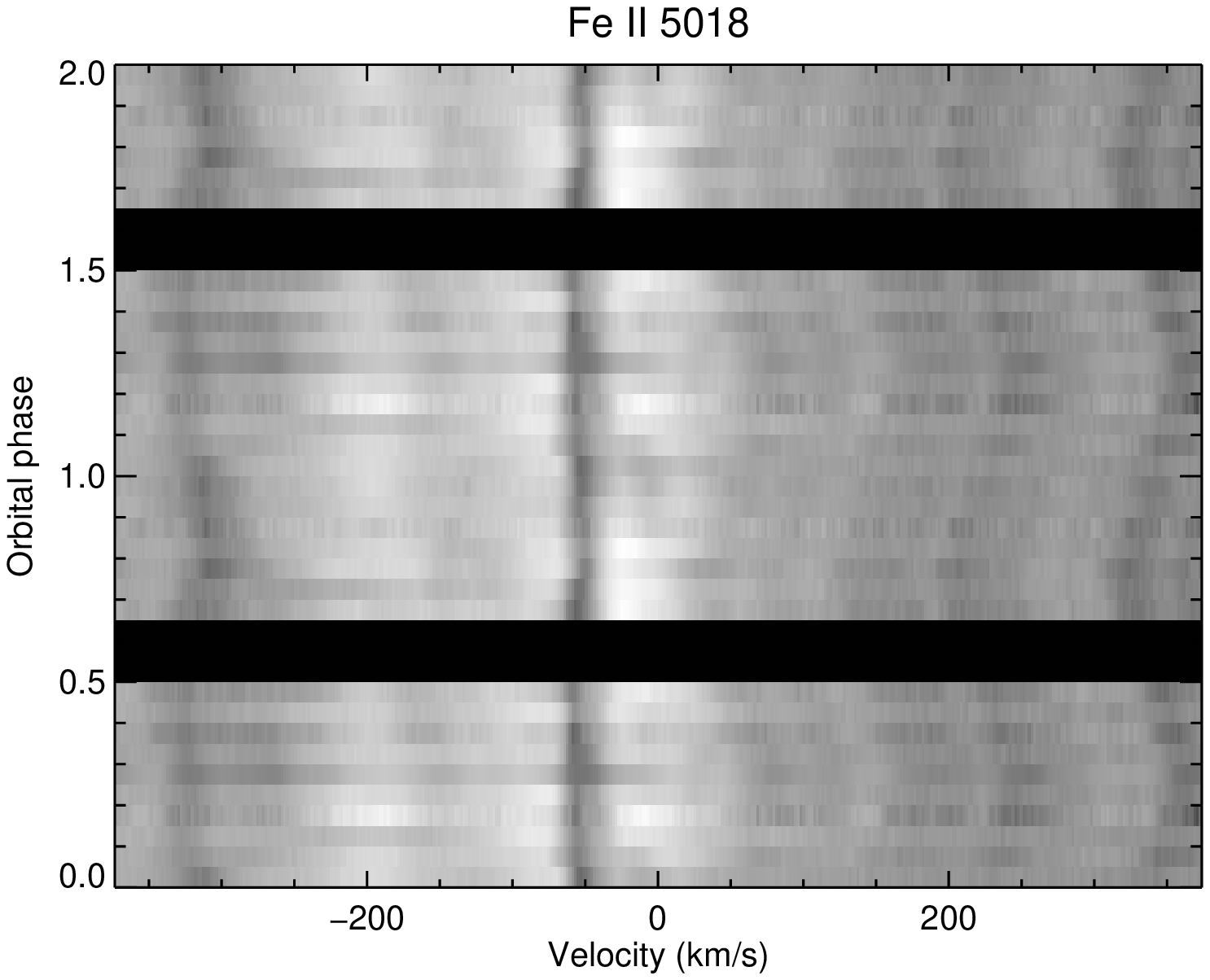}
\includegraphics[width=0.45\textwidth]{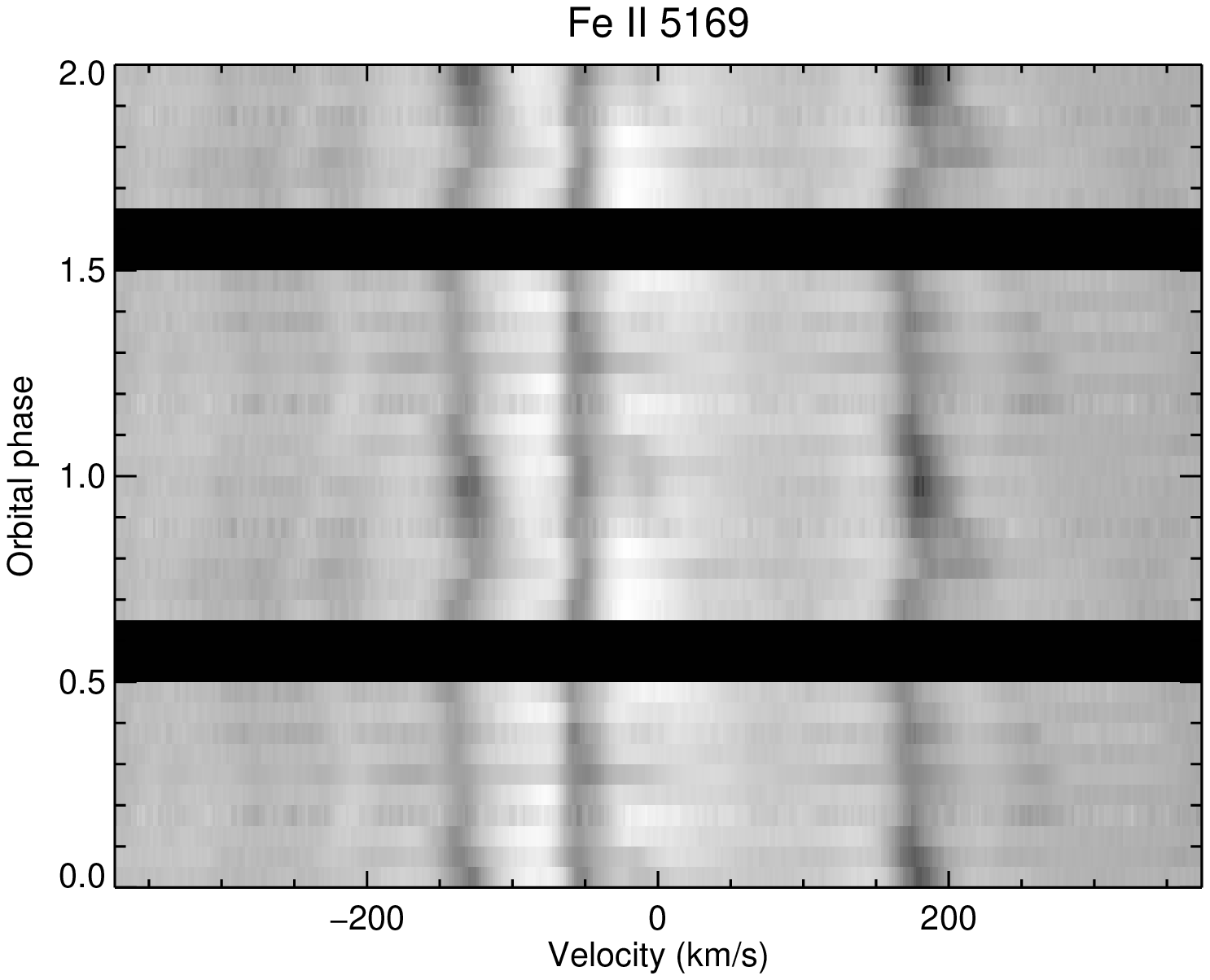}
\includegraphics[width=0.45\textwidth]{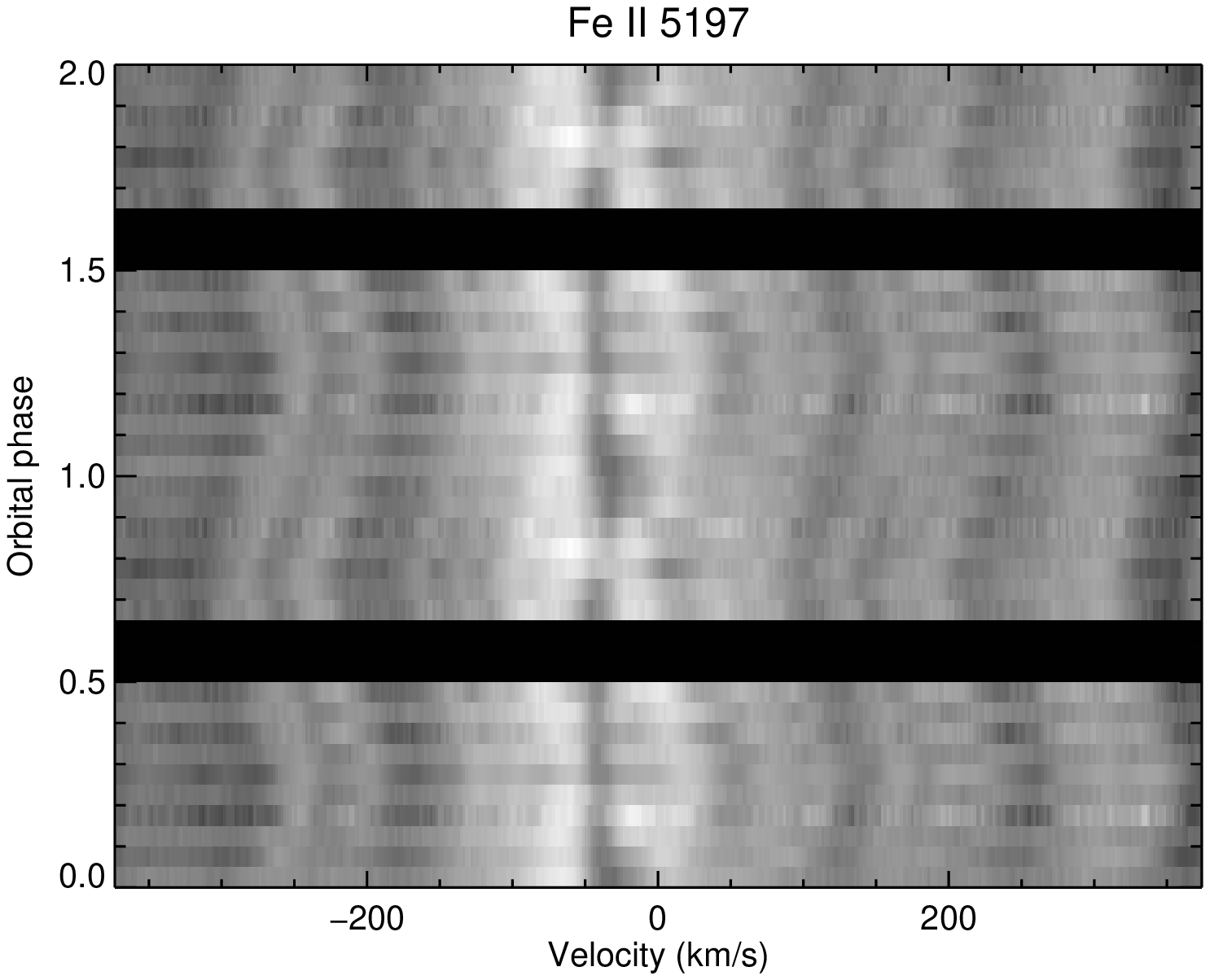}
\includegraphics[width=0.45\textwidth]{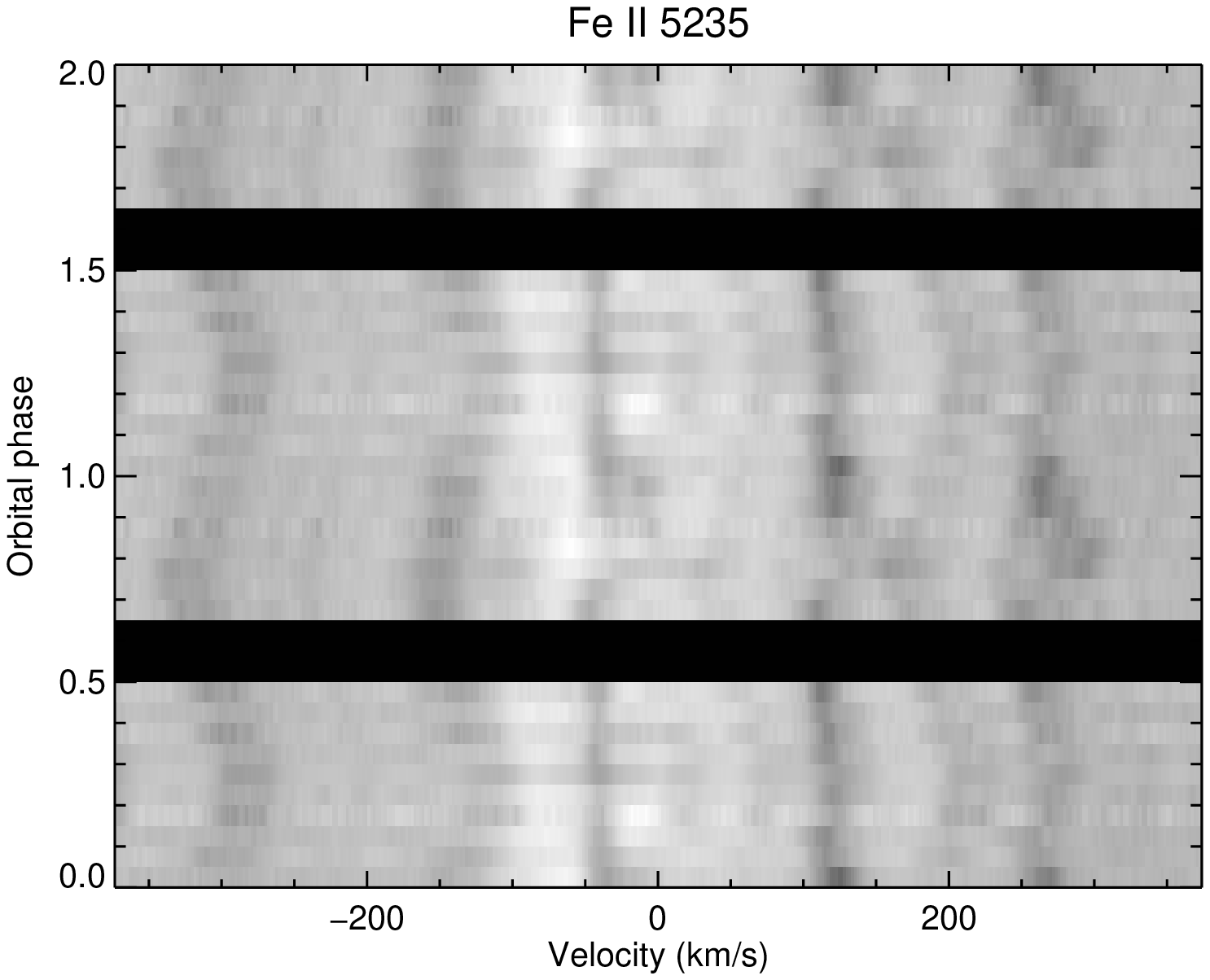}
\caption{The variability of Fe II 4233, 4924, 5018, 5169, 5197 and 5235 lines. Spectra have been normalised to continuum and phase binned into 20 bins and plotted over two orbital phases for clarity. The data has not been corrected for the systemic velocity (-35 km s$^{-1}$). Some faint absorption features from the red giant create the s-like curves.}
\label{pic:feii1seq}
\end{center}
\end{figure*}

\begin{figure*}
\begin{center}
\includegraphics[width=0.45\textwidth]{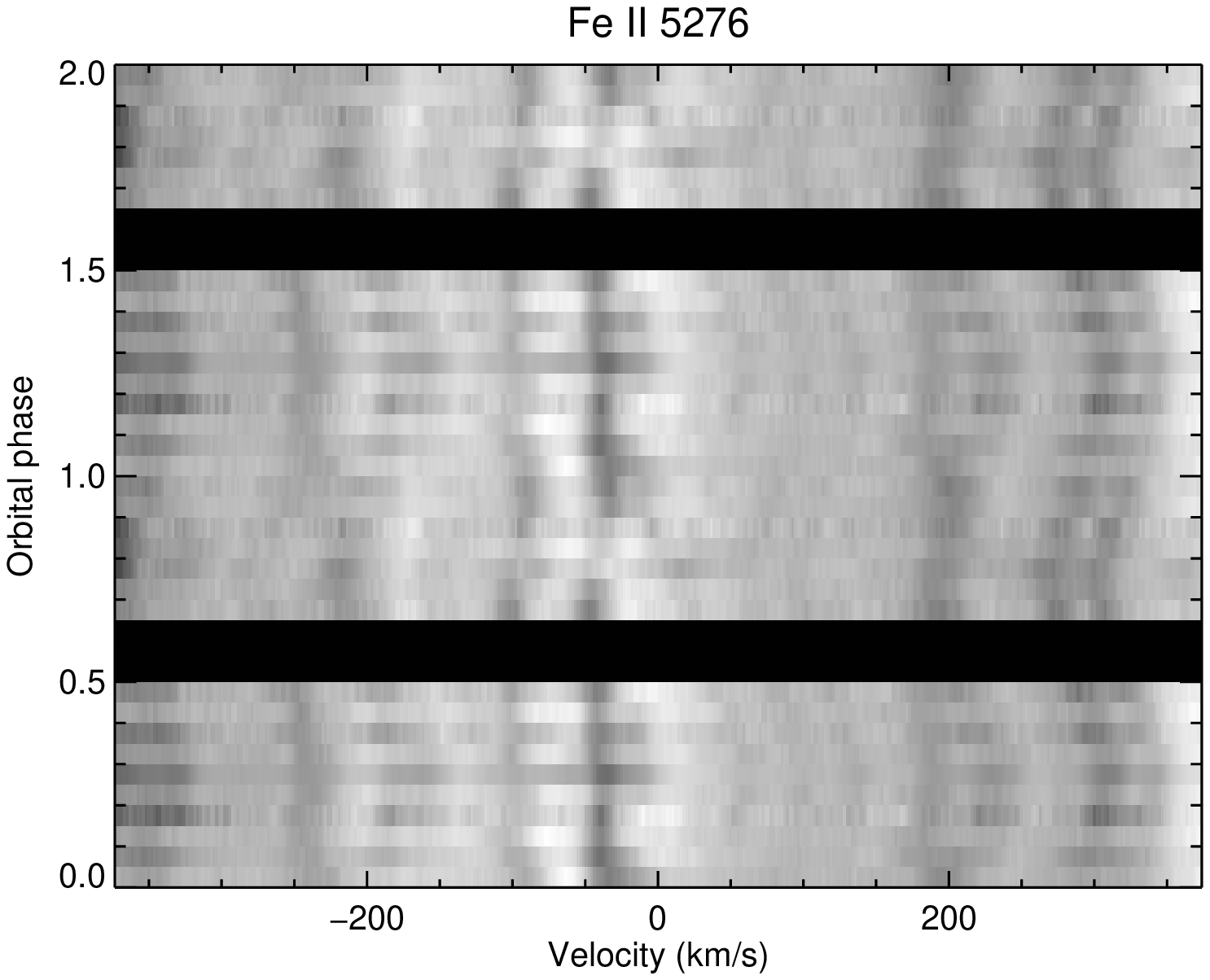}
 \includegraphics[width=0.45\textwidth]{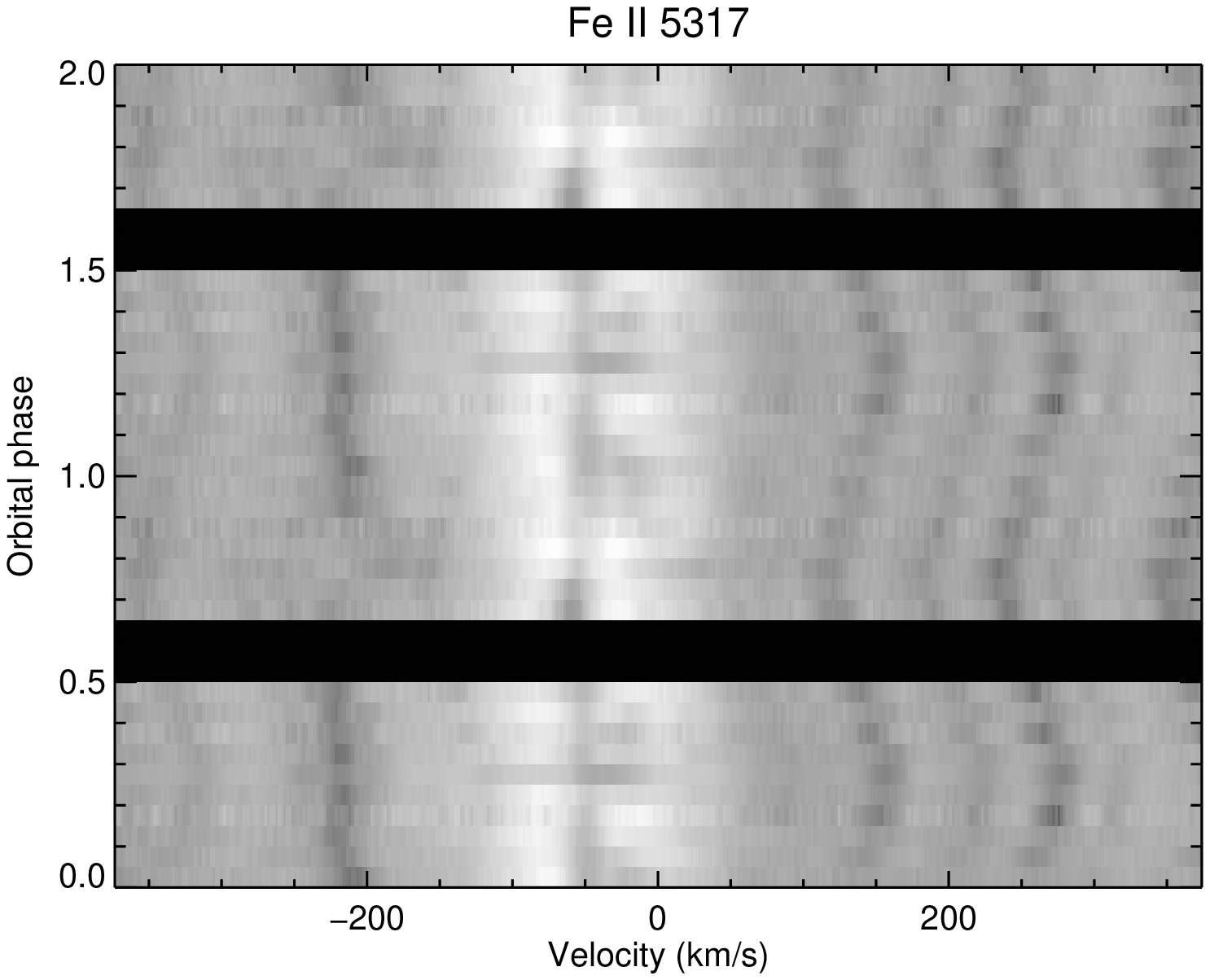}
\includegraphics[width=0.45\textwidth]{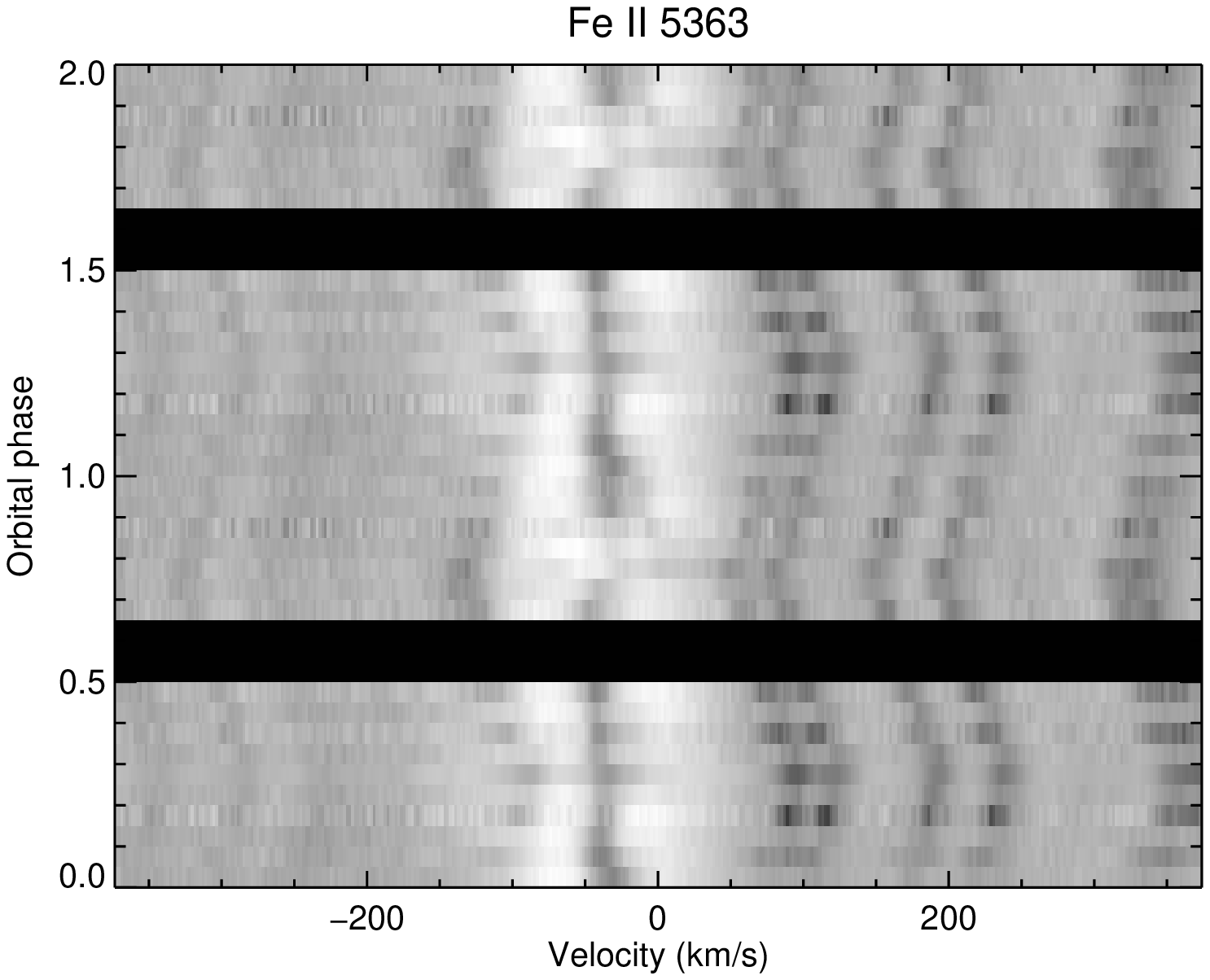}
\includegraphics[width=0.45\textwidth]{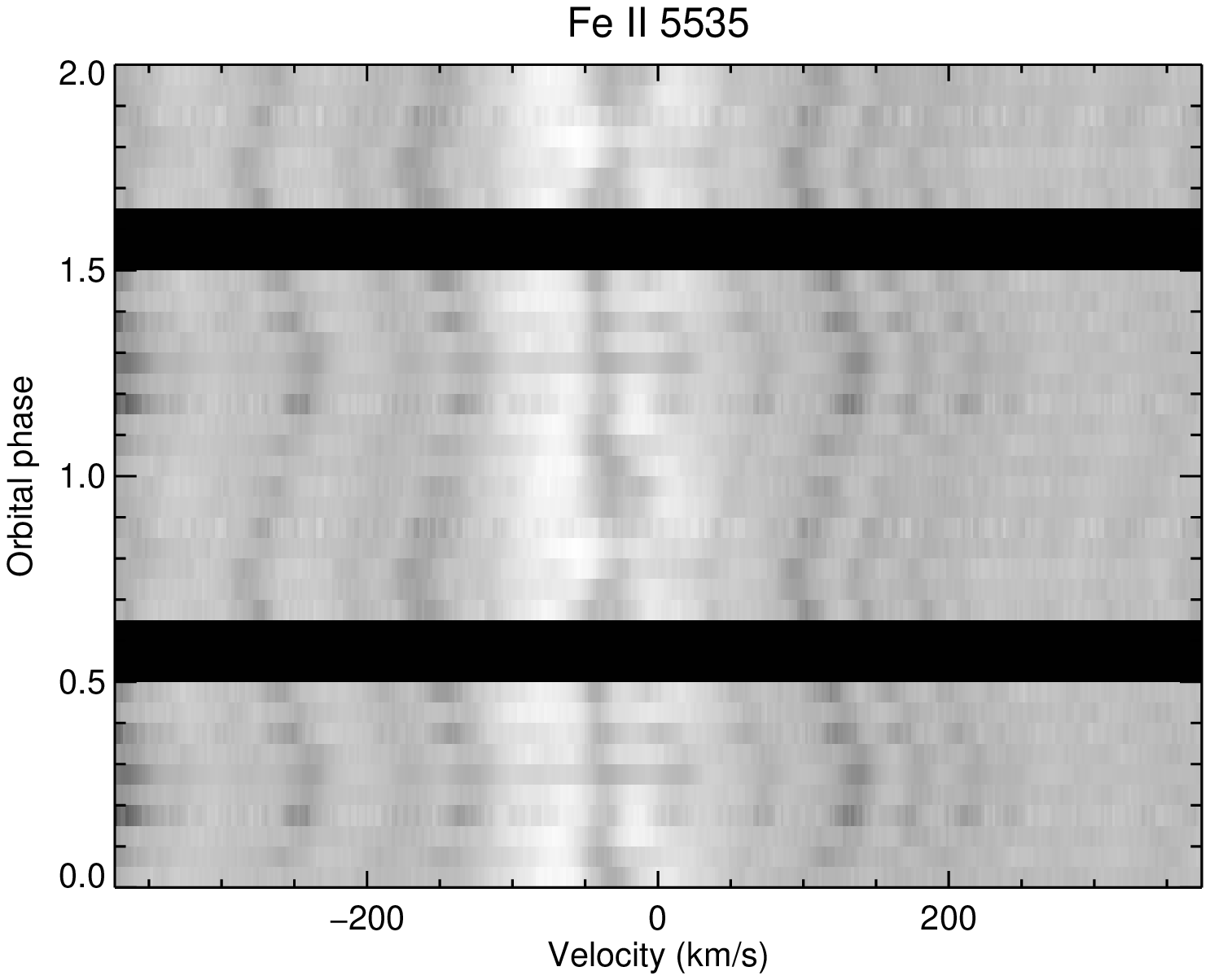}
\includegraphics[width=0.45\textwidth]{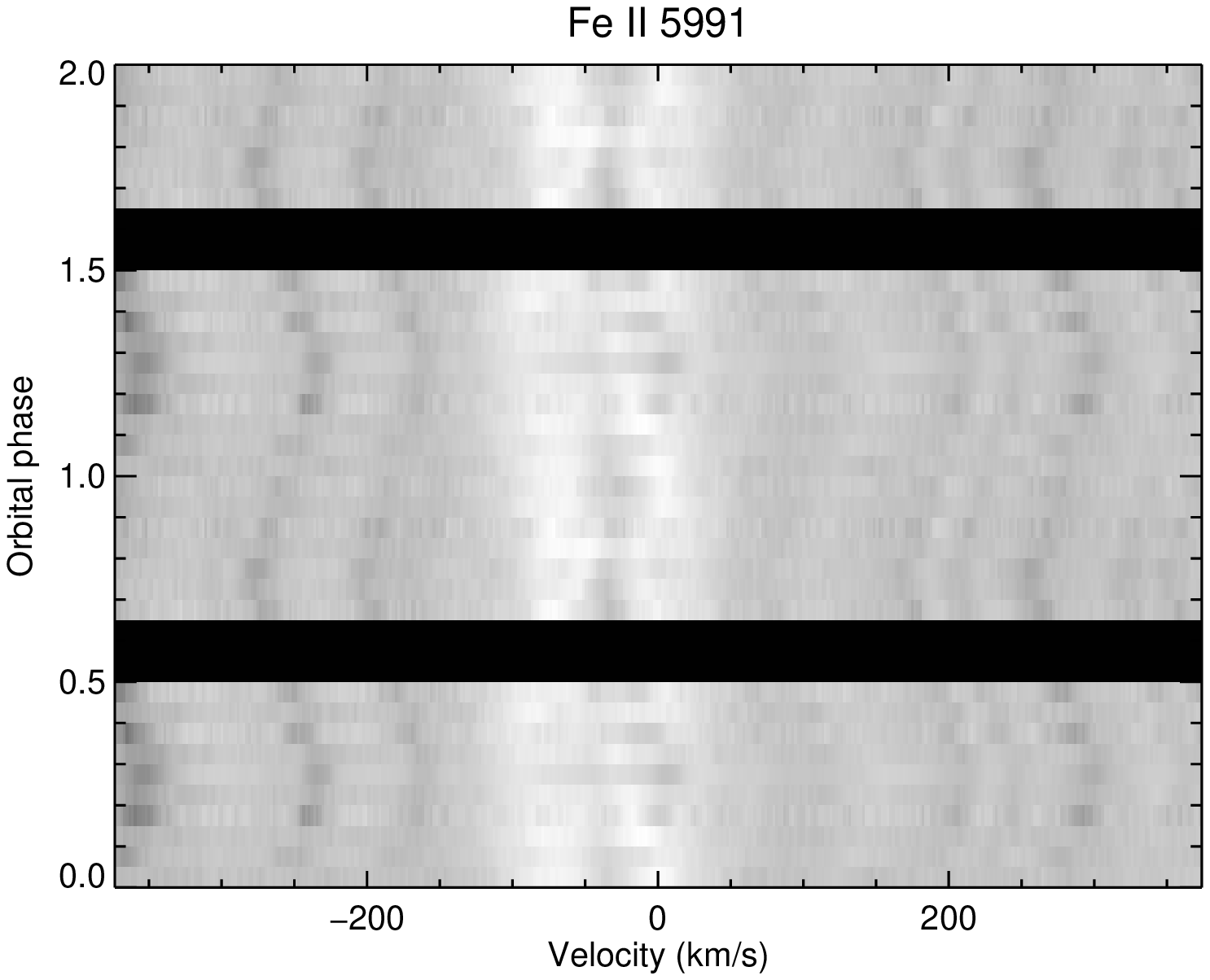}
\includegraphics[width=0.45\textwidth]{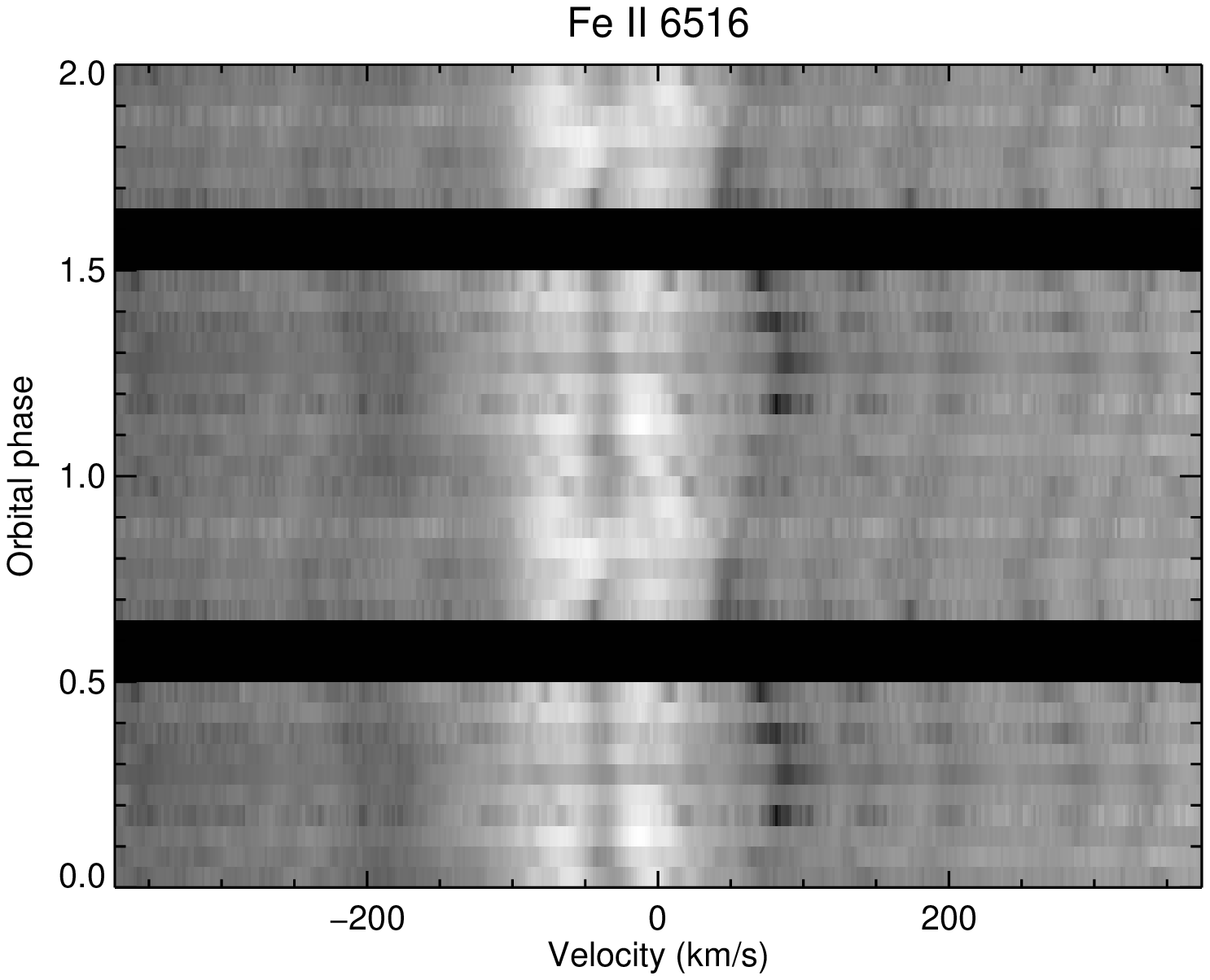}
\caption{The variability of Fe II 5276, 5317, 5363, 5535, 5991and 6516 lines. Spectra have been normalised to continuum and phase binned into 20 bins and plotted over two orbital phases for clarity. The data has not been corrected for the systemic velocity (-35 km s$^{-1}$). Some faint absorption features from the red giant cause the s-like curves.}
\label{pic:feii2seq}
\end{center}
\end{figure*}

\begin{figure}
\begin{center}
\includegraphics[width=0.45\textwidth, angle=0]{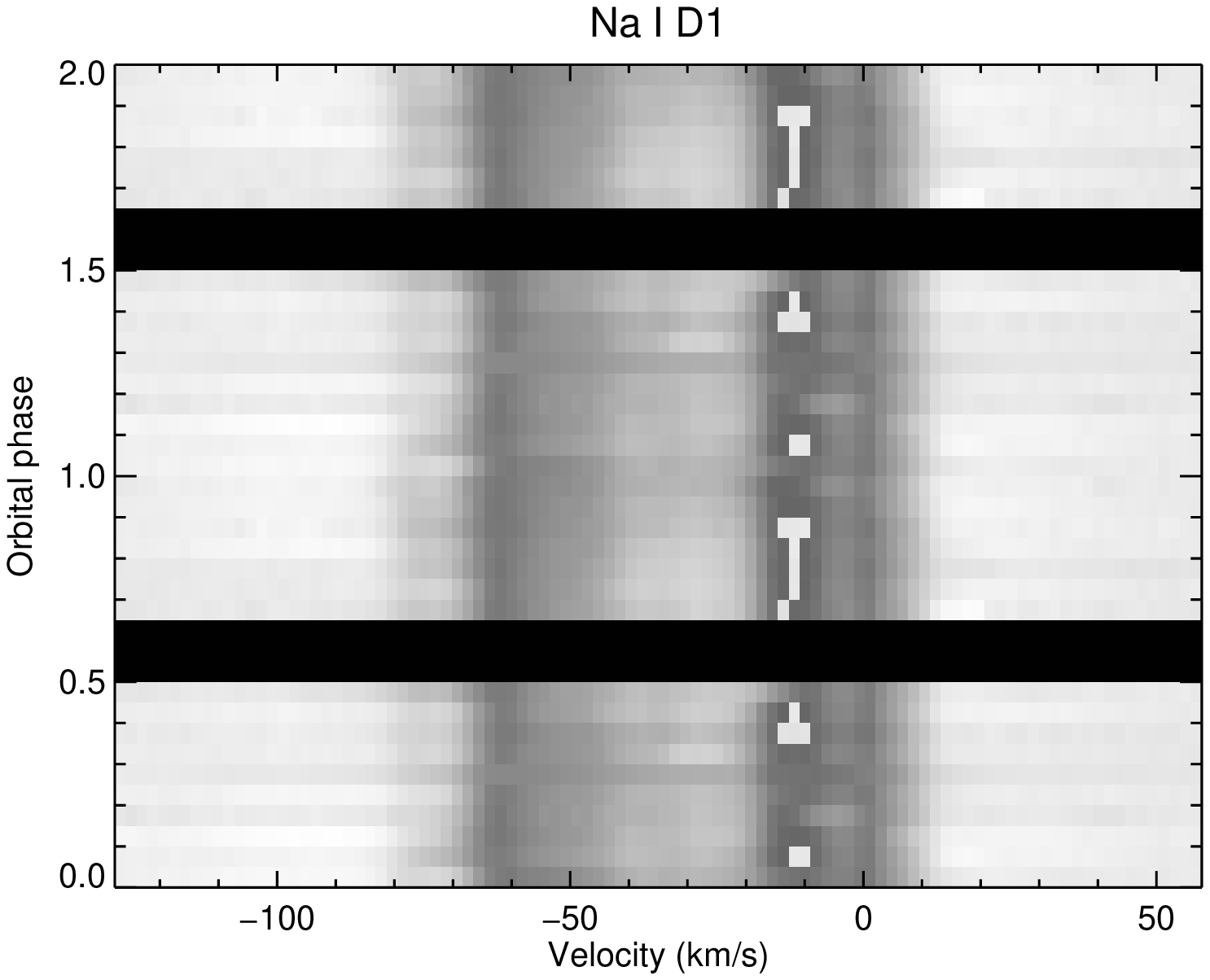}
\caption{The variability of Na I D1 line. Spectra have been normalised to continuum and phase binned into 20 bins and plotted over two orbital phases for clarity. The spectra have not been corrected for the systemic velocity.}
\label{pic:naid1seq}
\end{center}
\end{figure}


\bsp	
\label{lastpage}
\end{document}